\documentstyle[aas2pp4,twoside,aabib]{article}
\nonstopmode

\newcommand{\Jtco}{J=1$\rightarrow$0~$^{13}$CO}
\newcommand{\Jco}{J=1$\rightarrow$0~$^{12}$CO}

\def\Ma{{\cal{M}}_{ A}}
\def\Mai{{\cal{M}}_{A,i}}

\begin{document}
\title{A super-Alfv\'{e}nic model of dark clouds}

\author{Paolo Padoan}
\affil{Instituto Nacional de Astrof\'{i}sica, \'{O}ptica y Electr\'{o}nica, 
Apartado Postal 216, 72000 Puebla, M\'{e}xico}
\author{\AA ke Nordlund}
\affil{Astronomical Observatory and Theoretical Astrophysics Center, Juliane 
Maries Vej 30, DK-2100 Copenhagen, Denmark}
\authoremail{padoan@tac.dk}

\begin{abstract}

Supersonic random motions are observed in dark clouds and are traditionally 
interpreted as Alfv\'{e}n waves, but the possibility that these motions 
are super-Alfv\'{e}nic has not been ruled out. 

In this work we report the results of numerical experiments in two opposite 
regimes;
$\Ma\sim 1$ and $\Ma\gg 1$, where $\Ma$ is the initial Alfv\'{e}nic Mach
number --the ratio of the rms velocity to the Alfv\'{e}n speed.
Our results show that models with $\Ma\gg 1$ are consistent with the
observed properties of molecular clouds that 
we have tested --statistics of extinction measurements, Zeeman splitting
measurements of magnetic field strength, line width versus integrated antenna
temperature of molecular emission line spectra, statistical 
$B$--$n$ relation, and scatter in that relation-- while  
models with $\Ma \sim 1$ have properties that are in conflict 
with the observations.

We find that both the density and the magnetic field in molecular
clouds may be very intermittent.  The statistical distributions of magnetic 
field and gas density are related by a power law, with an index that decreases 
with time in experiments with decaying turbulence.  After about one dynamical time 
it stabilizes at $B\propto n^{0.4}$. Magnetically dominated cores form early 
in the evolution, while later on the intermittency in the density field wins out,
and also cores with weak field can be generated, by mass accretion along
magnetic field lines.

\end{abstract}

\keywords{
turbulence - ISM: kinematics and dynamics- magnetic field
}

\section{Introduction}

The observation of supersonic motions in molecular clouds (eg Zuckerman 
\nocite{Zuckerman+Palmer74}
\& Palmer 1974) raised the question of how these motions could be supported 
(Norman \& Silk 1980; Fleck 1981; Scalo \& Pumphrey 1982). 
\nocite{Norman+Silk80,Fleck81,Scalo+Pumphrey82}
Supersonic motions are expected to quickly dissipate their energy in highly 
radiative shocks, because of the very short cooling time of molecular gas or metal
rich atomic gas (Mestel \& Spitzer 1956; Spitzer 1968; Goldreich \& Kwan 1974). 
\nocite{Norman+Silk80,Fleck81,Scalo+Pumphrey82}

Strictly related was the issue of the support of molecular clouds (MCs) against
gravitational collapse, since it was soon realized that the observed motions
could not be understood as a gravitational collapse (Zuckerman \& Evans 1974; 
\nocite{Zuckerman+Evans74}
Morris et al.\ 1974), although MCs contain many Jeans' masses. 
\nocite{Morris+74}

\begin{sloppypar}
Theoreticians therefore formulated the hypothesis that MCs were primarily 
magnetically supported (Mestel 1965, Strittmatter 1966; Parker 1973;
\nocite{Mestel65,Strittmatter66,Parker73}
Mouschovias 1976a,b; McKee \& Zweibel 1995) and interpreted the observed 
\nocite{McKee+Zweibel95}
\nocite{Mouschovias76b}
\nocite{Mouschovias76a}
motions as long-wavelength hydro-magnetic waves (Arons \& Max 1975; Zweibel 
\nocite{Arons+Max75,Zweibel+Josafatsson83}
\& Josafatsson 1983; Elmegreen 1985, Falgarone \& Puget 1986). It was also shown
\nocite{Elmegreen85,Falgarone+Puget86}
that the properties of the observed `turbulence' (Larson 1981;
\nocite{Larson81}
Leung, Kutner \& Mead 1982; Myers 1983; Quiroga 1983; 
\nocite{Leung+82,Myers83,Quiroga83}
Sanders, Scoville \& Solomon 1985; 
\nocite{Sanders+85}
Goldsmith \& Arquilla 1985; Dame et al.\ 1986; Falgarone \& P\'{e}rault 1987) 
\nocite{Goldsmith+Arquilla85,Dame+86,Falgarone+Perault87}
could be understood if the motions were sub-Alfv\'{e}nic (Myers \& Goodman 1988;
\nocite{Myers+Goodman88}
Mouschovias \& Psaltis 1995, Xie 1997).
\nocite{Mouschovias+Psaltis95,Xie97}
\end{sloppypar}

Nevertheless, recent attempts to detect the Zeeman effect, in lines of molecules such 
as OH and CN, that probe regions of dense gas 
(Crutcher et al.\ 1993; Crutcher et al.\ 1996), 
\nocite{Crutcher+93,Crutcher+96}
resulted in a number of non-detections of the effect, and therefore in rather stringent upper 
limits for the magnetic field strength, despite the fact that regions expected to favor 
detections were targeted. It is therefore possible, that the field strength in dense 
molecular gas is weaker than assumed in theoretical studies.

\begin{sloppypar}
In the present work, we study the dynamics of models that share some 
properties with molecular clouds, using
numerical solutions of the equations of 3-D compressible magneto-hydrodynamics 
(MHD) in a regime of highly supersonic random motions. We idealize the 
system by omitting self gravity, ambipolar diffusion, radiative energy 
transfer, etc., and concentrate instead on the aspects that are 
directly related to the random, supersonic nature of the motions.
The main limitation of
our numerical models, compared with MCs, is the absence of gravity.
We have excluded gravity on purpose (our code is capable of handling
self-gravity), because one of the aims of our work
is to show that (magneto-)hydrodynamic processes alone are able to explain 
many of the observed properties of MCs.
Although gravity is certainly responsible for the final collapse of high density
regions into stars, we suspect that supersonic random motions are 
responsible for setting up many of the properties that characterize  
molecular clouds, with only minor contributions from gravity (Padoan 1995; 
\nocite{Padoan95}
Padoan, Jones \& Nordlund 1997; Padoan, Nordlund \& Jones 1997, 
\nocite{Padoan+97ext,Padoan+97imf}
\cite{Padoan+97cat}). 
The importance of supersonic motions in fragmenting the gas is certainly apparent
on small scales, where young and probably transient clumps are 
not originated by gravitational instability 
(Falgarone, Puget \& P\'{e}rault 1992; Langer et al. 1995).
\nocite{Falgarone+92,Langer+95}
\end{sloppypar}

Previous numerical studies have shown that compressible turbulence can
qualitatively explain several observational properties of MCs, even if
the effect of gravity and magnetic fields are not included.
The first two-dimensional (2-D) simulations of turbulence with rms Mach number
larger than one were performed by Passot \& Pouquet (1987). These were the
\nocite{Passot+Pouquet87}
first simulations where shocks were shown to develop inside a turbulent flow.
It was immediately recognized that shocks might have been responsible
for the fragmentation of the density field inside MCs, and especially
for the origin of their filamentary structure (Passot, Pouquet \& Woodward 1988).
\nocite{Passot+88}
The importance of shock formation inside mildly supersonic flows
was later confirmed in 3-D simulations by Lee, Lele \& Moin (1991).
\nocite{Lee+91}
Kimura \& Tosa (1993) simulated the passage of a strong shock through a
\nocite{Kimura+Tosa93}
turbulent molecular cloud, and found that this process can generate dense
clumps, with a power law mass spectrum. V\'{a}zquez-Semadeni (1994) made
\nocite{Vazquez-Semadeni94}
use of 2-D numerical simulations to show that supersonic turbulence
generates a very intermittent density field,
reminiscent of the clumpy nature of MCs. The density field was also found 
to be self-similar, which could be the reason for the hierarchical structure
of MCs (Scalo 1985, Houlahan \& Scalo 1992,
\nocite{Scalo85,Houlahan+Scalo92}
Elmegreen \& Falgarone 1996, Elmegreen 1997).
\nocite{Elmegreen+Falgarone96,Elmegreen97a}
Falgarone et al. (1994), analyzing the 
numerical simulation by Porter, Pouquet \& Woodward (1994),
argued that the properties 
\nocite{Porter+94}
of the profiles of molecular emission spectra from MCs can be interpreted as
arising from turbulent motions. 

Other numerical works included gravity in the turbulent flows, yet 
without describing the magnetic field. Turbulence was shown to be able
to prevent gravitational collapse (cf Chandrasekhar 1958; Arny 1971; 
\nocite{Chandrasekhar58,Arny71}
Bonazzola et al. 1987, 1992) in the 2-D numerical simulations by 
\nocite{Bonazzola+87,Bonazzola+92}
L\'{e}orat, Passot \& Pouquet (1990). 
\nocite{Leorat+90}
V\'{a}zquez-Semadeni, Passot \& Pouquet (1995) 
\nocite{Vazquez-Semadeni+95}
modeled the galactic disc, on the scale of 1 Kpc, as a 
turbulent self-gravitating flow. They simulate a two dimensional 
turbulent flow that is forced
by the energy released by star formation (expansions of HII regions), and found
that the main mechanism of cloud formation is the turbulent ram pressure,
rather than gravity. They were not able to form self gravitating clouds, due
to limitations in the thermal modeling, and the consequent low density 
contrast.

A three dimensional description of a magnetized self-gravitating cloud was
given by Carlberg \& Pudritz (1990), and was used to simulate 
\nocite{Carlberg+Pudritz90}
molecular emission spectra by Stenholm \& Pudritz (1993). 
\nocite{Stenholm+Pudritz93}
Carlberg \& Pudritz found that the magnetic field and hydro-magnetic
waves can support the cloud against gravity. The clouds contract,
because of ambipolar diffusion, on a time scale of approximately four 
free-fall times. These simulations do not solve the MHD equations, but
instead make use of a 'sticky particles' code. Energy is injected
in the form of a spectrum of Alfv\'{e}n waves, and the outcome of the
computation is dependent on the spectral index, that is a free parameter.
This way of forcing the particles is rather unphysical, because an arbitrary
spectrum of Alfv\'{e}n waves is imposed, instead of being obtained as a result
of the simulated magneto-hydrodynamics.
Passot, V\'{a}zquez-Semadeni \& Pouquet (1995)
\nocite{Passot+95}
introduced the magnetic field in their previous two dimensional model
for the galactic disc (V\'{a}zquez-Semadeni, Passot \& Pouquet 1995), 
\nocite{Vazquez-Semadeni+95}
and obtained a flow with rough equipartition of kinetic and magnetic energy,
probably in rough equipartition also with the mean thermal energy.
The same simulation, and others with larger density contrast and resolution,
have been studied by
V\'{a}zquez-Semadeni, Ballesteros-Paredes \& Rodr\'{i}guez (1997),
\nocite{Vazquez-Semadeni+97}
who were able to reproduce the observed relation between line-width
and size (Larson 1981). Gammie and Ostriker (1996) solved the MHD equations
\nocite{Larson81}
in a slab geometry, including self-gravity. By forcing the flow with
a nonlinear spectrum of MHD waves, they were able to prevent the 
gravitational collapse of their one dimensional cloud model.

Apart from the intentional exclusion of gravity, we differ from 
previous studies of turbulent flows in several respects. First of
all we have solved the MHD equations in three dimensions, while all
previous solutions are in two or one dimensions. In MHD the dimensionality
of the flow has a fundamental importance, because it determines the
topological freedom of the magnetic field. 
A price that has to be paid for using fully three-dimensional simulations
is a significant reduction of the number of points in each spatial
direction.  Thus, while our $128^3$ runs correspond to  $\sim 2\:10^6$
degrees of freedom, two dimensional simulation such as those of 
V\'azquez-Semadeni et al.\ (1997), that use up to 800 points in each 
direction, still need less than a third of the computing resources of our 
three dimensional simulations.  Studies of varying dimensionality are thus 
complementary, and may contribute to uncover different aspects of
the system under study.

An important property 
of our models is the high Mach numbers of the flows (rms up to $\sim$ 30, 
maximum $\sim$ 60), that allows us to investigate the super--sonic
regime of interstellar medium flows on rather large scales, up to about 
40~pc. Previous works have normally been limited to smaller
rms Mach number. For example some values of the initial rms Mach number
in decaying flows are: ${\cal M}_i=4$ in the two dimensional experiment
by Passot, Pouquet \& Woodward (1988); ${\cal M}_i=1$ in three dimensional 
runs by Porter, Pouquet \& Woodward 1992; ${\cal M}_i=10$ in one dimensional
simulations by Gammie \& Ostriker (1996); ${\cal M}_i=5$ in MacLow et al. 
(1998) in three dimensions. Other values of rms Mach number in driven numerical
experiments are: ${\cal M}=3$ in three dimensional simulations by
Lee, Lele \& Moin (1991); ${\cal M}=3$ in numerical experiments by 
Scalo et al. (1998); ${\cal M}=5$ in three dimensional simulations by Stone, 
Ostriker \& Gammie (1998). The two dimensional simulations by 
V\'{a}zquez--Semadeni, Passot \& Pouquet (1996) have large Mach numbers 
because of the cooling they include.
\nocite{Passot+88}
\nocite{MacLow_Puebla98}
\nocite{Porter+92}
\nocite{Scalo+98}
\nocite{Stone+98}
\nocite{Lee+91}
\nocite{Vazquez-Semadeni+96}

While previous models of
magnetized clouds focused on the role of the magnetic field as opposed to
gravity, and therefore have assumed a magnetic pressure much larger than the
gas pressure, the main purpose of the present work is to show that MCs are 
well described as flows with lower magnetic pressure than previously 
assumed. 

In this work we report the results of numerical experiments in two opposite 
regimes; $\Ma\gg 1$ and $\Ma\sim 1$, where $\Ma$ is the initial Alfv\'{e}nic Mach
number --the ratio of the rms velocity to the Alfv\'{e}n speed.
We argue that the observations 
of magnetic field strengths are consistent with a scenario where the mean 
magnetic pressure is dynamically low; i.e., where $\Ma^2 \gg 1$. 
Super--sonic and super--Alfv\'{e}nic magneto--hydrodynamic (MHD) flows 
develop a very intermittent spatial distribution of the magnetic energy, 
such that when the field is detected at a favorable position, its 
estimated strength is far higher than the mean field strength. 
Therefore, dense cores with sub-Alfv\'{e}nic velocity dispersion can 
be generated in super--Alfv\'{e}nic flows, in agreement with the observations.

\section{The experiments}

The study of the dynamics of MCs belongs to the field of random 
and super--sonic MHD flows. The Reynolds number and the magnetic 
Reynolds number in MC flows are very large. The random nature is therefore
a basic feature of the dynamics of these flows, and requires an appropriate 
description.

A realistic description of the dynamics
of molecular clouds have to be based on the numerical solution of
the compressible MHD equations in three dimensions, in a regime of
random and highly supersonic motions. We solve the compressible 
MHD equations in such a regime, at a numerical resolution of
$128^3$, with a code designed for turbulence and MHD turbulence experiments
(Nordlund, Galsgaard \& Stein 1994; Stein, Galsgaard \& Nordlund 1994; 
\nocite{Stein+94a}
\nocite{Nordlund+94a}
Galsgaard \& Nordlund 1996; Nordlund, Stein \& Galsgaard 1996;
Nordlund \& Galsgaard 1997),
\nocite{Nordlund+Galsgaard97mhd}
\nocite{Nordlund+96para95}
\nocite{Galsgaard+Nordlund95xc}
specifically adapted to be able to deal with very strong shocks and very large 
density contrasts. 

\nocite{Padoan+97ext}
\nocite{Padoan+97imf}

Although we have already developed a version of the code with the inclusion
of the gravitational force, all the experiments were run without gravity, 
for the following reasons:

\begin{itemize}

\item We are interested in studying if gas dynamics and 
magneto-hydrodynamics alone can shape distributions that are
similar to those observed, even without invoking gravity.

\item The observed motions have velocities comparable with the virial
velocity, or larger, on a range of scales, and the clouds are not 
free-falling.

\item If the results of our experiments are discussed only up to a time
shorter than or comparable to the dynamical  (or free-fall) time, all our
conclusions remain basically unchanged. This time is about a few million years on 
a scale of 10 pc, and clouds are not supposed to live much longer than that, 
before star formation takes place and becomes energetically important.

\end{itemize}

We note that the fact that clouds are observed to have velocities 
comparable to virial velocities, and are observed to not be 
free-falling, are not acceptable justifications for neglecting 
self-gravity; a complete model of interstellar clouds should instead
explain {\em why} they are not free-falling even in the presence of
gravity.  However, by concentrating first, in the present paper, on a situation
where gravity is neglected, we provide a point of reference for subsequent studies,
where we do include self-gravity (\cite{Padoan+97selfg}).
\nocite{Padoan+97selfg}

\subsection{The equations}

We solve the compressible MHD equations:

\def\vv{{\bf v}}
\def\jj{{\bf j}}
\def\bb{{\bf B}}
\def\lnr{\ln\rho}
\def\div{\nabla\cdot}

\begin{equation}
\label{0}
{\partial \ln\rho \over \partial t} + \vv \cdot \nabla\lnr = - \div \vv,
\end{equation}

  \begin{equation}
   {\partial{\vv} \over \partial t}
   + {\vv\cdot\nabla\vv}
  =
   - {P\over\rho} \nabla \ln P
   + {1\over\rho} {\jj} \times {\bb}
   + {\bf f},
  \label{1}
  \end{equation}

\begin{equation}
\label{4}
{\partial e \over \partial t} + {\vv} \cdot \nabla e = - {P \over \rho} \div 
{\vv} + Q_{\rm dissipation} + Q_{\rm radiation},
\end{equation}

\begin{equation}
{\partial{\bb} \over \partial t} = \nabla\times\vv\times\bb,
\label{2}
\end{equation}

\begin{equation}
\jj = \nabla\times\bb,
\label{3}
\end{equation}

\noindent
plus numerical diffusion terms, and
with periodic boundary conditions. $\vv$ is the velocity, $\bb$ the
magnetic field, ${\bf f}$ an external force ($=0$ in these particular
experiments), and $P = \rho T$ is the pressure at $T \approx$ const. 

Conditions in the cold molecular clouds that we are modeling are such that
an isothermal approximation is adequate; the radiative heat exchange is
so efficient that the temperature remains low in most places.  Even if
the temperature momentarily increases in shocks, the subsequent cooling is
rapid, and the result is shock structures that are qualitatively and
quantitatively similar to isothermal shocks.

We have thus used isothermal conditions in most of our runs and
have verified that this is appropriate, by
rerunning segments of some experiments using the full energy equation.
No significant change of the statistics was found and, since using the
full energy equation increases the cost of the experiments considerably
(the strong cooling required to maintain a low temperature forces a
much smaller time step), we performed most of the experiments at
constant temperature.

The absence of an explicit resistivity $\eta$ in the induction equation 
corresponds to an assumption of flux freezing on well resolved scales.
The code uses shock and current sheet capturing techniques to ensure
that magnetic and viscous dissipation at the smallest resolved scales 
provide the necessary dissipation paths for magnetic and kinetic energy.
As shown by Galsgaard \& Nordlund (1996, 1997), dissipation of 
\nocite{Galsgaard+Nordlund95xc,Galsgaard+Nordlund95ff}
magnetic energy in highly turbulent, compressible MHD 
plasmas occurs at a rate that
is independent of the details of the small scale dissipation. 
In ordinary hydrodynamic turbulence the corresponding property is
one of the cornerstones of Kolmogorov (1941) scaling.
\nocite{Kolmogorov41}

The works by Galsgaard \& Nordlund (1996, 1997) refer to 
compressible flows, but since the ratio of mean kinetic energy
to mean magnetic energy is small, motions are in practice 
roughly incompressible, at least in the directions perpendicular
to the local magnetic field.  In the case referred to in the
present paper as Model A, the mean kinetic energy is instead large
compared to the mean magnetic energy, and the situation might
be rather different, and more similar to Burger's turbulence
than both classical Kolmogorov turbulence, and the situation studied 
by Galsgaard and Nordlund.  However, Burger's turbulence is also
characterized by having a dissipation that is, in the limit,
independent of the value of the viscosity (cf.\ Saffman 1968,
p. 485, sec. 6).

We have not included ambipolar diffusion in any of the driven experiments.
The value of the rms Mach number in the driven experiments is comparable
to the value estimated in molecular clouds on a scale range from 3~pc 
to 50~pc. On such scales, ambipolar diffusion occurs in a time significantly 
longer than the dynamical time, as recently shown by Myers \& Khersonsky 
(1995) and 
\nocite{Myers+Khersonsky95}
is expected to be of secondary importance (see also
Elmegreen \& Fiebig 1993).
\nocite{Elmegreen+Fiebig93}

On sufficiently small scales and high densities, ambipolar diffusion
can have significant effects
(Zweibel \& Brandenburg 1994, 1997; Tagger et al\. 1995).
\nocite{Tagger+95,Zweibel+Brandenburg94,Zweibel+Brandenburg97}
Since the decay runs have lower rms Mach number (smaller scale),
and since ambipolar diffusion is considered an important
dissipation mechanism in magnetized clouds (McKee et al. 1993), 
the ambipolar drift has been included in some of the decaying runs.
\nocite{McKee+93}

\subsection{The code}

The code solves the compressible MHD equations on a 3D staggered mesh, with volume
centered mass density and thermal energy, face centered velocity and
magnetic field components, and edge centered electric currents and electric fields
(Nordlund, Stein \& Galsgaard 1996).
\nocite{Nordlund+96para95}

The original code works with ``per-unit-volume'' variables; mass density,
momenta, and thermal energy per unit volume.  In the super-sonic regime relevant 
in
the present application, we found it advantageous to rewrite the code in
terms of ``per-unit-mass'' variables; $\ln\rho$, $u$, and 
$E=\frac{3}{2}\frac{P}{\rho}$.
With these variables, the time evolution of all variables is
governed by equations of the type
\begin{equation}
{D f \over Dt} = {\partial f \over \partial t} + \vv \cdot \nabla f = ... ;
\end{equation}
i.e., equations that specify the time rate of change following the motion.
These are better conditioned than the divergence type equations that
result from using per-unit-volume variables (the large---order $M^2$---
density variation in isothermal shocks cause the per-unit-volume fluxes to
vary over several orders of magnitude).

We use spatial derivatives accurate to
6th order, interpolation accurate to 5th order, and Hyman's 3rd order time 
stepping
method (Hyman 1979).
\nocite{Hyman1979}

In order to minimize the viscous and resistive influence on well resolved
scales, we use monotonic 3rd order hyper-diffusive fluxes instead of normal
diffusive fluxes, and in order to capture hydrodynamic and
magneto-hydrodynamic shocks we add diffusivities proportional to the
negative part of the velocity divergence, and resistivity proportional to
the negative part of the cross-field (two-dimensional) velocity divergence.
Further details of the numerical methods are given by
Nordlund, Stein \& Galsgaard (1996) and Nordlund \& Galsgaard (1997).
\nocite{Nordlund+Galsgaard97mhd}
\nocite{Nordlund+96para95}

\subsection{Weak and strong magnetic field}

For the purpose of this work we have run several experiments,
that can be divided into two main groups: i) super--Alfv\'{e}nic
runs (models A), with $\Mai\approx10$ (in one case $\Mai\approx30$), 
and $\beta_{i} \ge 1$; ii) equipartition (of kinetic and magnetic 
energies) runs (models B), with $\Mai\approx1$ and $\beta_{i} \ll 1$;
where $\Mai$ is the initial value of the Alfv\'{e}nic Mach number
(the rms velocity of the flow divided by the Alfv\'{e}n velocity)
and $\beta_{i}=(P_{g}/P_{m})_{i}$ is the initial ratio
of gas pressure to magnetic pressure. Notice that the magnetic
energy grows considerably in the beginning of the super--Alfv\'{e}nic
runs, and therefore the value of $\Ma$ can quickly decrease by a factor 
of two and the value of $\beta$ by a factor of four.

Both models A and B are divided in two sub--groups: decaying runs
(simply A and B), and randomly driven runs (Ad and Bd). Models
of the same type can also differ by the initial rms Mach number 
of the flow (the rms velocity of the flow divided by the speed of 
sound), or by the inclusion or not of ambipolar diffusion.
A list of the models is given in Table~1, together with the initial
values of their parameters.

In all experiments the initial density is uniform, and the initial
velocity is random. We generate the velocity field in Fourier space,
and we give power, with a normal distribution, only to the Fourier
components in the shell of wave-numbers $1 \leq k L / 2 \pi \leq 2$. We perform
a Helmholtz decomposition, and use only the solenoidal 
component of the initial velocity. The initial magnetic field is 
uniform, and is oriented parallel to the $z$ axis: 
${\mathbf B}=B_0{\mathbf \hat{z}}$.

\section{Results}

In this section some properties of the models that may be related 
to observational quantities are discussed. 
The initial values of some physical parameters of these models are listed 
in Table~2. There are only two numerical parameters in the models: 
${\cal M}$ and $\Ma$.
In order to rescale the models to physical units, we use the following 
empirical Larson's relations (Larson 1981):\nocite{Larson81}

\begin{equation}
{\cal{M}}=4.0\left(\frac{L}{1pc}\right)^{0.5}
\end{equation} 
where a temperature $T=10$~K is assumed, and
\begin{equation}
\langle n \rangle=2.0\times10^3\left(\frac{L}{1pc}\right)^{-1}
\end{equation} 
that is equivalent to a constant mean surface density.
With these two relations, the value of ${\cal{M}}$
determines the value of $L$, the linear size of 
the model cloud, and of $\langle n \rangle$, the average number density.
The physical unit of velocity in the code
is the isothermal speed of sound, $C_s$, and the physical unit of the
magnetic field is $C_s(4\pi\langle\rho\rangle)^{\frac{1}{2}}$ (cgs).

\subsection{Dissipation of supersonic motions}

Molecular cloud lifetimes are estimated to be of the order 
of a few dynamical times, while super--sonic motions should 
dissipate their kinetic energy in shocks in about one dynamical time.  
One of the original motivations for models of magnetized clouds with 
equipartition of kinetic, magnetic and gravitational energies
was the belief that MHD waves decay on a longer time--scale than
random super--sonic and super--Alfv\'{e}nic flows, and therefore 
could offer an explanation for the long lifetime of molecular clouds.
Zweibel \& Josafatsson (1983) and Elmegreen (1985),
\nocite{Elmegreen85}
\nocite{Zweibel+Josafatsson83}
noticed that decay of MHD waves can be rather short, but
their studies are based on the ambipolar diffusion dissipation.
Ambipolar diffusion is in fact believed to be the most important 
dissipation mechanism in magnetized clouds (\cite{McKee+93}).

We have run decaying experiments both with and without ambipolar
diffusion, from which we can see that the inclusion of ambipolar diffusion
does not change very much the decay time--scale. In our decaying 
equipartition models, the most important energy dissipation occurs
because motions along the magnetic field lines produce strong shocks, just
as in super--Alfv\'{e}nic models.  This was found also by Gammie and Ostriker 
(1996). 
\nocite{Gammie+Ostriker96}
The ratio of the decay time to the instantaneous dynamical time is plotted 
in Fig.~\ref{fig1} for models without ambipolar diffusion, and in
Fig.~\ref{fig2} for similar models, but with the inclusion of ambipolar diffusion.
Decay times of kinetic, magnetic and total energies are plotted versus
the time in units of the initial dynamical time, $\tau_{dyn}=L/\sigma_{v}$. 
After a short initial transient (of the order of $0.5 \tau_{dyn}$ because
the initial velocity field has power up to $k L / 2 \pi = 2$),
the decay time is in all cases comparable to the instantaneous dynamical
time. Thus, the strength of the magnetic field, and also the importance
of ambipolar diffusion, do not significantly affect the time--scale of 
decay of the flow.  

A simple interpretation of the insensitivity of the level of dissipation 
to details of the dissipation mechanism is that it is related to the origin
of the scaling behavior of turbulent flows.  We observe a similar behavior
in our super--sonic experiments as in experiments with sub--sonic turbulence;
a power law energy power spectrum is established already after
a fraction of a dynamical time. The scaling is such that the decay
of energy does not depend on the particular dissipation process that occurs
on the small viscous and resistive scales, but is rather determined by the 
large scale kinematics, characterized by the dynamical time--scale. 

It may be concluded that equipartition models do not offer any advantage
as far as the energy decay is concerned, compared with super--Alfv\'{e}nic 
models. \citetext{MacLow_Puebla98} and \citetext{Stone+98} have recently
shown that the decay time in equipartition models can be even shorter than
in super--Alfv\'{e}nic models. The problem of the lifetime of molecular
clouds is not to be solved by the intervention of a strong magnetic field,
but more likely by continuous driving of the turbulence.
Cascading energy from distant supernova activity is a likely source of
galactic turbulence (\cite{Korpi+98a,Korpi+98b,Korpi+99a}). 
Also, bipolar outflows from stars formed inside molecular clouds can 
provide significant amounts of energy 
(Reipurth, Bally \& Devine, 1997).  Moreover, cloud life times
are not very long (Blitz \& Shu 1980), and could perhaps even be comparable
with their dynamical time (Elmegreen \& Efremov, 1998). 
\nocite{Blitz+Shu80}
\nocite{Elmegreen+Efremov98}
\nocite{Reipurth+97}

In this connection it should be pointed out that, even though energy 
sources {\em on the average} replenish turbulence in clouds, the driving 
may be intermittent in time as well as in space, and turbulence in 
individual clouds may thus, at times, be decaying.

\subsection{Probability Distribution of Magnetic Field Strength}

The main difference between the super--Alfv\'{e}nic model and the equipartition 
model is the fact that in the latter the kinetic energy is not sufficiently large
to compress the magnetic field significantly, while in the former
very strong density enhancements produce large local enhancement of the
magnetic field strength. Moreover, expansions generate large voids
with magnetic field strength much smaller than the mean field in the
super--Alfv\'{e}nic model, but not in the equipartition model.

As a result, runs of type A produce very intermittent probability
distributions of the field strength, with a roughly exponential tail.
Models of type B, instead, are characterized by approximately Gaussian
probability distributions of the field strength. In Fig.~\ref{fig3},
the field strength distributions from the last snapshots of model Ad2 
(thick line) and model Bd1 (thin line) are plotted. In model Ad2,
$\langle B \rangle=4.5$~$\mu$G, but values of field strength up to 100~$\mu$G
are reached in high density regions. The values of $B$ span almost four
orders of magnitude, 0.03~$\mu$G$<B<$100~$\mu$G, in model Ad2, 
while they span only one order of magnitude in model Bd1, 
4.5~$\mu$G$<B<$46~$\mu$G.

In the super--Alfv\'{e}nic model most of the mass concentrates in 
regions of large gas density (see below), where the field strength 
is also large. Because of this, approximately $0.5\%$ of the total 
mass of the system contains a field 10 times stronger than the mean value. 
For example, the last snapshot of model Ad2 corresponds to molecular clouds 
on the scale of 15~pc, with $\langle n \rangle=133$~H$_2$cm$^{-3}$. 
It may be taken, therefore,
as a model for a molecular cloud of about $2\times10^4$~M$_{\odot}$.
Since $\langle B \rangle=4.5$~$\mu$G, one clump of about 100~M$_{\odot}$ can be found
in the cloud, with $B\sim 45$~$\mu$G, according to model Ad2.

The examples show that the formation of cores with rather large field 
strength is predicted in the super--Alfv\'{e}nic model, even if the
mean magnetic field strength in the cloud is quite small, and comparable
with the mean galactic field strength. As a result, Zeeman splitting 
observations should not easily detect the magnetic field in clouds,
apart from inside particularly favorable high density clumps.
Conversely, if the magnetic field is detected in a cloud core,
it is likely that the mean field in the cloud is much smaller
than the estimated field strength in that core, if the cloud dynamics is 
described by a random super--sonic and super--Alfv\'{e}nic flow,
as in model Ad2.

\subsection{Probability Distribution of Gas Density: Stellar Extinction}

In a highly radiative gas, the most important effect of supersonic random 
flows is the fragmentation of the medium into filaments and clumps with 
very large density contrast. Supersonic turbulence can indeed be the 
reason for the observed fragmentation of molecular clouds (eg Padoan \& Nordlund 1998).
\nocite{Padoan+Nordlund_Puebla98}
It is interesting therefore to quantify the statistics of the density
distribution that arise from supersonic turbulence. V\'{a}zquez--Semadeni
(1994) found that the density distribution in his two dimensional numerical 
simulations was consistent with a Log--Normal. Padoan, Nordlund \& Jones (1997)
also found density distributions consistent with a Log--Normal in their
three dimensional numerical simulations, while Scalo et al. (1998) studied the effect of 
a polytropic equation of state on the density distribution. Nordlund \& 
Padoan (1998) used a new set of numerical experiments to show that 
\nocite{Nordlund+Padoan98puebla}
even polytropic flows produce a density distribution similar to the Log-Normal,
but slightly skewed. They also argue that such a distribution is due to the
fact that the dynamics does not depend on the mean density.

Observationally it is difficult to measure the gas volume density 
in molecular clouds, but stellar extinction measurements can provide
information about the projected gas density in clouds. 
Lada et al. (1994) suggested the use of stellar extinction measurements as 
a test for models of the structure of molecular clouds. Padoan, Jones, 
\& Nordlund (1997) have shown that near-infrared
\nocite{Padoan+97ext}
stellar extinction measurements can be used to 
infer the three dimensional probability distribution of the gas density 
in dark clouds. They have shown that there is qualitative and quantitative
agreement between the inferred properties of the three dimensional density 
distribution in the dark cloud IC5146 (Lada et al. 1994), and the 
properties of the three dimensional density distribution in their 
experiments of random supersonic flows. 

The comparison between numerical models and observational data
is based on the plot of the dispersion of the extinction measurements
in cells, versus the mean extinction in the same cells (Lada et al.\ 1994). 
\nocite{Lada+94}
In Padoan, Jones \& Nordlund (1997), the theoretical 
\nocite{Padoan+97ext}
plots are calculated starting from random density fields of given statistics 
and power spectra. Here we generate the same plot, but calculated 
directly from the density field of the experiments. A random distribution 
of stars is generated and it is assumed that most stars are in the background
(that is the model cloud is between the stars and the observer). 
The visual extinction is proportional to the column density of the cloud 
sampled by the line--of--sight to any given star:

\begin{equation}
A_V/mag=\frac{N(H+H_2)}{2\times10^{21}cm^{-2}}
\end{equation}
(Bohlin et al. 1978) \nocite{Bohlin+78}. 
A uniform grid is used to subdivide the whole area into cells, each containing 
an average number of five stars, as in Lada et al. (1994). The mean extinction 
$A_V$ and its dispersion $\sigma(A_V)$ are measured inside each cell, and
$\sigma(A_V)$ is plotted against $A_V$. 

We have performed this test with all our models. In Fig.~\ref{fig4} we show the results
for the models Ad2 and Bd1. These two models both have very large rms Mach
number and differ from each other because model Ad2 is super--Alfv\'{e}nic,
while model Bd1 has approximate equipartition of magnetic and kinetic energy. 
We use the last snapshots from the two runs, where the values of
Mach number and Alfv\'{e}nic Mach number are ${\cal M}\approx 15$ and 
${\cal M}_A\approx 5$ respectively, for model Ad2,
and ${\cal M}\approx 12$ and ${\cal M}_A\approx 1$ for model Bd1.
According to the Larson's relations we use in the present work to rescale
our numerical models to physical units (Eqs. (7) and (8)), these snapshots
correspond to molecular clouds on the scale of 14~pc (Ad2) and 9~pc (Bd1).
Using the data by Lada et al. (1994) and by Dobashi et al. (1992), one may infer
an rms Mach number ${\cal M}\approx 10$ for the dark cloud IC5146, comparable to the 
Mach number values in the simulations. However, the particular scale is not important
as for the mean or total column density of model or real clouds. This is because the
Larson's relations are such that the column density of
clouds is independent of size. The scale is instead important for 
determining the slope of the upper envelope of the $A_V$--$\sigma(A_V)$ plot. 
As shown in Padoan, Jones \& Nordlund (1997), that slope is roughly proportional
to the standard deviation of the three dimensional density distribution, which is
also proportional to the Mach number of the flow. So the slope is expected
to be larger on larger scales (higher rms Mach number).

It is clear for Fig.~\ref{fig4} that the super--Alfv\'{e}nic model reproduces very well 
the observed plot, while the equipartition model (Bd4) does not. 
Although model Bd4 has an rms Mach number comparable to 
the one in the cloud IC5146, the values of $A_V$ and $\sigma(A_V)$ in the model
span a much smaller range than in the observations. This is also illustrated by the 
histogram of extinction (column density), shown on the left panels of Fig.~\ref{fig4}.
The equipartition model behaves like an elastic medium in the directions perpendicular
to the magnetic field lines, and this reduces very much its ability to generate
a sufficiently complex density field. In particular, while compressions on all three
dimensions can generate a topology with large holes and adjacent lines--of--sights
with very large fluctuations in extinction (model Ad2), compressions in only one 
dimension (model Bd1) can do that only to a smaller extent.

\subsection{Synthetic spectra: line width versus integrated temperature}

Most of the data from dark clouds consist of molecular emission line
observations. The intensity of a particular emission line depends 
on many factors such as gas density, kinetic temperature,
radiation field, and chemical abundances. Moreover, the line
profiles are strongly dependent on the gas kinematics, but in a 
rather complex way, because only radial velocities are available,
and both line widths and intensities are quantities projected along
the line--of--sight. It is very difficult to extract reliable
information about the three dimensional structure and kinematics
of molecular clouds directly from observational data, and
making use of simple models assuming local thermodynamic equilibrium (LTE),
smooth velocity fields, and oversimplified density distribution.
Part of the problem is due to the complexity of the cloud structure 
and of the internal random motions in clouds.

\citetext{Padoan+97cat} have recently calculated maps of synthetic
molecular spectra that may be used to infer intrinsic properties of
molecular clouds by comparison with observed spectra. Details of
the method can be found in \citetext{Padoan+97cat}. Here we simply 
remark that the synthetic spectra are obtained with a non-LTE
Monte Carlo radiative transfer code (Juvela 1997), and the radiative
transfer calculations are performed on a cloud model that is similar
to the ones used in
present work. Images of synthetic molecular maps of different
transitions and molecules are shown in \citetext{Padoan+97cat}, together with
statistics of the spectral line profiles. In that work, a super--Alfv\'{e}nic
flow was used, and was shown to reproduce observed properties of clouds
(see also \cite{Padoan+98per}). In the present work we are interested 
also in the comparison between the super--Alfv\'{e}nic and equipartition
models, and we have therefore solved the radiative transfer problem using 
cloud models Ad2 and Bd4 (the radiative transfer calculations have been 
kindly provided by Mika Juvela). Notice that
although the initial Mach number of model Ad2 is almost 30, it later decreases
(due to its initial conditions and not strong enough random driving), to a value
comparable to the one in model Bd4. We use the last snapshots from each model,
which correspond to the physical regime in typical molecular clouds on the 
scale of about 10--15~pc. 

Comparisons of molecular spectra between the two models have shown that
while the two cases are rather similar in many of their observable 
properties, they are very different in the relation between 
\Jtco\ line width and \Jtco\ integrated antenna temperature. The integrated
antenna temperature is roughly proportional to the gas column density
(cf.\ Dickman 1978, and \cite{Padoan+97co}, for a critical discussion based on non-LTE
radiative transfer), and its histogram is shown in Fig.~\ref{fig5}.
\nocite{Dickman78}
\nocite{Padoan+98per}
The equipartition model has a narrower distribution than the super--Alfv\'{e}nic 
model, consistent with the discussion of dust extinction in the previous subsection.
The histogram of line width is also narrower for model Bd4 that for model
Ad2 (not shown), although the spectra averaged over all synthetic
maps are almost identical (same line width). The relation between the line width
and the integrated antenna temperature is also quite different in the two 
cases. The equipartition model shows almost no growth of line width with increasing 
\Jtco\ antenna temperature, while in model Ad2 the growth is 
very significant, as shown in Fig.\ref{fig6}. 

Heyer, Carpenter \& Ladd (1996) find that in the star forming 
giant molecular clouds they observed
\nocite{Heyer+96}
the \Jco\ line width grows with \Jtco\ integrated antenna temperature 
(roughly proportional to column density),
and discuss the possibility of using this property to infer the dynamic 
importance of the magnetic field in the cloud motions. They argue that, 
if the motions observed in molecular clouds were Alfv\'{e}n waves, 
like in theoretical models of dark clouds (for example in Carlberg \& Pudritz 1990), 
then the line width should decrease with column density. Based on this, they conclude
that the motions in the clouds they observed are not consistent with Alfv\'{e}n waves.
We find that the \Jtco\ line width in the equipartition model is in general almost 
independent of column density (\Jtco\ integrated antenna temperature), and in some 
cases even decreases with column density, as predicted by Heyer, Carpenter and 
Ladd (1996) for the \Jco\ line width. The equipartition model is therefore in 
conflict with the observations, while the super--Alfv\'{e}nic model shows a relation
between line width and column density that is similar to the observational result. 

In order to better constrain the importance of the magnetic field in the dynamics
of molecular cloud internal motions, it would be useful to measure line width 
versus integrated temperature in many different clouds. We expect that the 
magnetic field could play an important role in some cores with large field 
strength, and this could appear in the relation between \Jtco\ line width 
and column density. Cores with strong magnetic field (predicted in the present 
super--Alfv\'{e}nic scenario for molecular cloud dynamics) have already been 
identified by OH Zeeman splitting measurements, and should be the 
target of such a study.

\subsection{The $B$--$n$ relation}  

As discussed above, the equipartition model generates a smaller
density contrast than the super--Alfv\'{e}nic model, and also 
generates much smaller 
fluctuations in the magnetic field strength. The relation between 
field strength and gas density ($B$--$n$ relation) is therefore expected 
to be rather different in the two models. The $B$--$n$ relation in the models  
may be compared with observational results.

In regions of maser emission, at densities of about $n=10^7cm^{-3}$, 
a field strength of the order of $B=10^3-10^4\mu G$ is observed, 
while in regions of molecular emission, with approximately $n=10^2-10^3 cm^{-3}$, 
the field is found to be of the order of $B=10\mu G$ (eg Myers \& Goodman 1988).
\nocite{Myers+Goodman88}
A relation of the type $B\propto n^{0.3-0.6}$ may be deduced
from the observations (eg Troland, Crutcher \& Kaz\`{e}s 1986; Heiles 1987; 
\nocite{Troland+86,Heiles87}
Dudorov 1991), 
\nocite{Dudorov91}
but it is quite uncertain, especially in the light of the 
above discussion about the intermittency of the distribution of the magnetic 
energy.

We have updated the $B$--$n$ relation by Troland \& Heiles (1986) with 
more recent measurements: the HI Zeeman observations by 
Verschuur (1995); the OH and CN Zeeman splitting measurements by Crutcher 
et al. (1993, 1996, 1998); the OH maser measurements by Johnston et al. (1989); 
the H$_2$O maser measurements by Fiebig \& Gusten (1989). 

In Fig.~\ref{fig7}, observational detections
of the field--strength and upper limits are plotted versus the estimated value
of the gas density. Although most of the attempts
to detect the Zeeman effect result in non-detections, and therefore in upper
limits to the field--strength, there are some detections of $B$ also in the
range of density found in molecular clouds. $B$ is of course detected preferentially
in the regions where it is particularly strong, which are hardly representative
of the typical physical conditions inside molecular clouds. Fig.~\ref{fig7}
shows a clear power--law relation between $B$ and $n$, defined by the regions
with strongest magnetic field. At the same time, there are many estimated upper limits
(over 100 listed in publications and only summarized in Fig.~\ref{fig7}) that very likely
probe many regions with values of $B$ weaker than the ones detected in regions
of similar densities. The observations therefore clearly indicate that the 
$B-n$ relation has a large scatter, although a power-law upper envelope 
is well defined over approximately 10 orders of magnitude in gas density
and 5 orders of magnitude in field--strength. The upper envelope is roughly

\begin{equation}
B\approx100\mu G\left(\frac{n}{10^3cm^{-3}}\right)^{0.4} ,
\end{equation}
and the scatter at $n=10^3$~cm$^{-3}$ is at least two orders of magnitude,
in the approximate range $1-100$~$\mu$G.

A super-Alfv\'{e}nic flow (eg model Ad2) naturally develops a $B-n$ relation with a power
law upper envelope and a large scatter, and also provides a large density
contrast at the same time. The $B$--$n$ relation is initially almost linear
(see below for an explanation), and flattens with time. The upper envelope 
of the relation also flattens with time, and after about one dynamical time 
it stabilizes to $B\propto n^{0.4}$, as in the observations. In Fig.~\ref{fig7} 
contour lines obtained from the $B$--$n$ scatter plot in model Ad2 (thick lines)
are shown, together with the observational data. Model Ad2 corresponds to typical 
molecular clouds on the scale of 20~pc (rms Mach number $\approx 20$). The 
particular snapshot used in Fig.~\ref{fig7} has ${\cal M}=15$ 
and $\langle B \rangle=4.5$~$\mu$G.
Although the mean field strength is so low (of the order of the mean galactic 
field), the fluctuations of field strength are every large ($0.03<B<100$~$\mu$G),
and the largest values of the magnetic field strength (upper envelope) at each density
are very close to the observational values. In model Ad2 there is a clear power--law 
upper envelope over about 4 orders of magnitude in density and almost 2 in $B$,
which matches the observations nicely. The scatter in the $B$--$n$ relation
of model Ad2 is also consistent with the observations. In order to achieve the
densest values of field strength estimated in molecular clouds using CN Zeeman 
splitting and masers, higher numerical resolution would be necessary, or 
perhaps self--gravity should be included in the calculations.

The result for an equipartition model (Bd1) is also overplotted in Fig.~\ref{fig7}
(thin lines). The model is inconsistent with the observations in two ways:
i) the ranges of density and field strength values are too small (see the two
subsections above) and ii) neither the upper envelope nor the mean field
strength at each density define a $B$--$n$ relation similar to the observational
one. In particular, most observational upper limits on the field strength, 
and a few detections, are completely inconsistent with the equipartition model.

While super--sonic and super--Alfv\'{e}nic turbulence naturally explains
the results of Zeeman splitting measurements, including the $B$--$n$ relation
and its scatter, such results are not easily explained by the alternative
equipartition model.  

The lack of correlation between $B$ and $n$ in model Bd1 is due to the fact
that most of the density enhancement occurs by convergent flows 
along the field lines, and these motions do not affect the magnetic field. 

One could possibly argue that the lack of correlation of $B$ and $n$ in model
Bd1 is consistent with the majority of observations, because often no field is 
detected, and hence nothing can be said about the correlation.  However, as further
discussed in the next Section, the upper limits of the non-detections speak
against this interpretation.  Also, the cases where a field {\em 
is} detected would then have to be explained with {\em ad hoc} arguments,
rather than as a natural part of a statistical distribution.

\subsection{Cloud and flow topology}

An understanding of the spatial structure of the density field in dark clouds
is very important for a correct interpretation of observational data,
and for the formulation of a number of physical models.
In this subsection we provide our interpretation of the dynamical phenomena
occurring in the simulations, based on extensive three dimensional browsing of the data,
and illustrated here by images of the density field and plots of statistical
correlations of the velocity and magnetic fields.

The snapshots in Fig.~\ref{fig8a} and Fig.~\ref{fig8b} 
give an idea of the dimensionality of the structures in the density field.
The topology of the density field 
in experiment Ad2 (Fig.~\ref{fig8a})
has a clear evolution in time. In the very beginning, until 
$t\approx0.6t_{dyn,0}$, the density grows predominantly in sheets. 
These are the fronts of blobs of coherent motion, advancing at 
supersonic velocity. Later, these fronts start to intersect 
each other, and the density increases especially in filaments 
(at the intersections of fronts). The evolution continues with the
intersection of filaments into knots, at $t\approx1.5t_{dyn,0}$. The 
fully developed topology, at $t=2.0t_{dyn}$, is characterized by both 
filaments and knots (cores). 
Also in experiment Bd1 (Fig.~\ref{fig8b}) 
the density grows initially in sheets, but the transition from sheets to
filaments is mainly due to motions along field lines, rather than to 
intersection of sheets. In the equipartition model Bd1, in fact, density 
enhancements cannot move freely in the three space directions, but only 
along magnetic field lines. It is therefore typical that density enhancements
such as sheets are torn apart, by motions along field lines, into filaments 
approximately aligned with the
magnetic field. The approximate alignment of the filaments is visible
in the left panel of Fig.~\ref{fig8b}. Later in time, the density structure
becomes more complex (right panel of Fig.~\ref{fig8b}), but still shows 
some evidence of the alignment of the density filaments. 

The evolution of the magnitudes of the mass density and the magnetic field
may be discussed with reference to Lagrangian version of the continuity equation,
\begin{equation}
\label{Lagrrho}
{{\rm D} \ln \rho \over {\rm D} t} = - \div \vv,
\end{equation}
and the scalar induction equation
\begin{equation}
\label{LagrB}
{{\rm D} \ln |B| \over {\rm D} t} = -\nabla\cdot \vv_{\perp} 
+ \hat{B}\cdot(\hat{B}\cdot\nabla)\vv_{\perp} ,
\end{equation}
where $\vv_{\perp}\cdot$ is the velocity perpendicular to the
magnetic field, and $-\nabla\cdot \vv_{\perp}$ usually is
the dominant term on the RHS.

Although $-\div\vv$ vanishes for the solenoidal initial condition, the
supersonic motions rapidly lead to the formation of shock fronts, where
the local value of $-\div\vv$ is large and positive because of the 
discontinuity in the velocity perpendicular to the shock.

The initially homogeneous magnetic field is carried along by the 
perpendicular components of the velocity field, and is hence also
collected into sheets, except at those rare locations where the initial
field happens to be strictly parallel to the velocity field.  This explains
why sheets initially form in both mass density and magnetic flux
density, and why the $B$--$n$ relation initially has an exponent close to unity.

Note that the usual argument for a slope of 2/3, that applies to isotropic
and non-shocking compressible motion does not apply here, because of the
development of discontinuities. In term of Eqs.\ \ref{Lagrrho} and
\ref{LagrB}, the 2/3 follows if $-\nabla\cdot \vv_{\perp}$ typically picks
up two of three statistically equivalent contributions to $-\div\vv$. 
However, at a
shock, the divergence is dominated by the derivative in one particular
direction; the one perpendicular to the shock front. The magnetic field
that is swept into the discontinuity quickly becomes almost parallel to the
shock front, because the component in the plane of the shock grows
exponentially with time. Thus, as long as the topology is dominated by
sheets, the mass density and the magnetic flux grow more or less in unison
in the sheets, corresponding to an exponent in the $B-n$ relation close to
unity.

In the subsequent evolution, there are effects that tend to reduce the
exponent in the $B$--$n$ relation.   First,  the non-linear evolution of the
initially solenoidal velocity field also leads to the development of 
regions of space with a positive divergence, in which both the mass 
density and the magnetic flux density decline.  In these regions, there
is no particular dominance of the cross-field divergence,  and thus the
three-dimensional divergence picks up an additional contribution relative
to the two-dimensional divergence, consistent with the classical 2/3
argument outlined above.

In experiment B1 the motions are mainly Alfv\'{e}n waves, and therefore the
velocity is predominantly perpendicular to the direction of the magnetic
field. This is illustrated in Fig.~\ref{fig9}, where we have plotted the histogram
of $cos(\alpha)$, where $\alpha$ is the angle between ${\bf v}$ and ${\bf
B}$.   Note that, even though motions across the field lines are the 
most common ones, it is the less frequent motions along the field 
lines that dominate the dissipation.  The motions across field lines are 
subject to magnetic restoring forces, and do not lead to substantial density
enhancements.  

In experiment A1, the magnetic field is advected by the flow, and the
stretching of field lines instead produces some alignment between ${\bf v}$
and ${\bf B}$, already before one dynamical time has passed, as illustrated in
Fig.~\ref{fig9}.

Alignment between $\bb$ and $\vv$ may be caused by two, complementary
effects: 1) {\em Dynamical alignment} is expected when the magnetic energy
approaches and exceeds the kinetic energy; the Lorentz force then forces
the flow to be predominantly along the magnetic field lines. 2) {\em
Kinematic alignment} occurs when a spatially non-uniform velocity field
causes stretching of magnetic field lines, and hence a correlation of $\bb$
and $\vv$. Pure shear, for example, tends to align an embedded magnetic
field with the direction of the flow. 

Motions that are aligned with the magnetic field affect the mass density
without affecting the magnetic flux density. In particular, the non-linear
concentration into first sheets and then filaments due to the interaction
of shock fronts continues into the formation of knots in the density field,
by the convergence of matter flowing along filaments. There is no
corresponding process available to a divergence free vector field such as
the magnetic field; once the field has concentrated into filaments, it
cannot concentrate further; the magnetic field in a filament is insensitive
to flow along the filament.

In the same way that converging flows along the magnetic field may lead to
extreme concentrations of mass, those regions where the flow is diverging
along magnetic field lines may lead to extreme rarefactions of mass,
without affecting the magnetic flux density. In $B$--$n$ scatter plots, this
corresponds to the development of more extreme excursions of the mass,
relative to those of the magnetic flux density, and hence a flattening of
the $B$--$n$ relation with time.

In model A1, dynamical alignment is at most significant in the few cores that
develop a strong (sub-Alfv\'{e}nic) magnetic field in the early evolution
of the experiment. In scatter plots of $B$ against $n$ most contributions
come from regions where dynamical alignment is unimportant. We thus
conclude that the evolution of the $B-n$ relation in model A, towards a
smaller exponent with time, is caused by the kinematic alignment of $\bb$
and $\vv$.

\section{Discussion}

It is difficult to get an objective view on the magnetic 
field strength in dark clouds from the literature. The reasons are the following:

\begin{itemize}

\item Negative results from observational programs in which detections 
have been reported are likely to remain unpublished.

\item The positions searched for Zeeman splitting never represent a
statistically meaningful sample. Favorable regions are always selected,
because the observations are very time consuming.

\item The total number of regions in dark clouds, for which OH Zeeman 
observations are published, is still small.

\end{itemize} 

Although all Zeeman splitting measurements are biased towards 
regions where a strong field is expected (high density cores), 
in many cases the field is not detected, and upper limits are quite 
stringent. As an example, Crutcher et al.\ (1993) selected 
four cores in the Taurus dark cloud 
\nocite{Crutcher+93}
complex, two in the Libra complex, two in $\rho$ Oph, one in the Orion molecular
ridge, one position in L889, and the core of B1 (Barnard 1), in the Perseus region.
The only certain detection is in
the cloud B1. For the other regions, the weighted average value of the field is
$+2.7\pm 1.5 \mu G$ in Taurus, $-2.1\pm 2.8 \mu G$ in Libra, $+6.8\pm 2.5\mu G$
in $\rho$Oph, $-0.6\pm2.1\mu G$ in L889, and $-4.7\pm3.5 \mu G$ in L1647.
CN Zeeman splitting measurements have also provided rather low upper limits 
to the field strength, compared with estimated equipartition values
(Crutcher et al. 1996), and also magnetic field detections well
below equipartition (Crutcher et al. 1998).
\nocite{Crutcher+96}
\nocite{Crutcher_Puebla98}
A rather low average field of $+9\mu G$ was also found in Cas A by 
Heiles and Stevens (1986).
\nocite{Heiles+Stevens86}

Orion A and Orion B are instead two examples of detections of intense 
magnetic field. OH Zeeman splitting in Orion A revealed a magnetic field 
strength of $B=-125\mu G$ (Troland, Crutcher, \& Kaz\`{e}s 1986), and 
\nocite{Troland+86} $B=+38\mu G$ toward Orion B (Crutcher \& Kaz\`{e}s 1983).
\nocite{Crutcher+Kazes83}

The OH Zeeman splitting should probe regions with $n=10^3-10^4$ cm$^{-3}$,
while the CN Zeeman splitting, should probe regions of $n\approx10^6$ cm$^{-3}$.

In our model Ad2, the mean field strength is 
$B=4.5\mu$~G, and the mean density $\langle n\rangle=1.3\times 10^2$~cm$^{-3}$;
$20\%$ of the total mass is in dense clumps, 
with density ten times larger than the mean, $n\approx3\times 10^3$~cm$^{-3}$, 
and field strength five times, or more, larger than the mean, $B\ge25\mu$~G. 
Therefore, even if field strengths of about $40\mu G$ are detected 
sometimes in dense cores, the mean Alfv\'{e}n velocity in the molecular cloud
may be just comparable to the sound speed, $\langle v_{A}^2\rangle^{1/2}\approx C_{S}$.

It may be concluded that model Ad2 is consistent with the observational
estimates of magnetic field strength in dark clouds.  The particular values
of the magnetic field used in the comparison here should not be taken too
literally; the small scale field strengths could be larger than estimated by the
Zeeman effect if the magnetic field is tangled (but notice that in Fig.~\ref{fig7} the
observational measurements have been multiplied by a factor of two, to account
for the random orientation of the magnetic field lines relative to the 
line--of--sight). 

It is difficult to envisage how the measurements could be consistent with
equipartition models such as Bd1 or Bd4, however, since the kinetic 
energy does not exceed the magnetic energy. As demonstrated by
Galsgaard \& Nordlund (1996), a magnetically dominated plasma is able to
\nocite{Galsgaard+Nordlund95xc}
quickly dissipate structural complexity, independent of the value of the
resistivity. Thus, a sub--Alfv\'{e}nic field could not remain strongly
tangled, and hence could not avoid detection. Using the same argument, we
expect those cores where a strong (sub--Alfv\'{e}nic) field has indeed been
observed to have a relatively simple magnetic field structure.

Model Ad2 can be used for the description of a typical molecular cloud with
linear size of about $15$~pc and $\langle n\rangle=130$~cm$^{-3}$. At the time
$t=1.0t_{dyn}$, the relation $B$--$n$ is then:

\begin{equation}
B\approx4.5\mu G(\frac{n}{100 cm^{-3}})^{0.7}
\end{equation}
Since the exponent is $>0.5$, most of the dense cores 
are found to have magnetic pressure larger than thermal pressure, at early times.
The lowering with time of the exponent in the $B$--$n$ power law means that 
later on, the dominance of the magnetic field in dense cores tends to be
reduced. Although the flow is random and approximately isotropic, the 
kinematic alignment of ${\bf B}$ with ${\bf v}$ makes dense cores accrete 
mass along ${\bf B}$ at an increased rate. Therefore, the accretion of mass
around dense cores, embedded in a random flow with $\beta\approx 1$, is such that,
while magnetic pressure becomes dominant over thermal pressure during the initial
phase of turbulent fragmentation, later on, in some cores, magnetic pressure 
can decreases to the level of thermal pressure, even in the absence of gravity, 
and on a time scale that is competitive with ambipolar diffusion. This 
mechanism could be relevant for the process of star formation.

A central point that we want to stress in this paper is that the fact of finding
some cloud cores, with sub--Alfv\'{e}nic velocity dispersions and with magnetic 
pressure larger than thermal pressure, does not
necessarily mean that the dynamics of molecular clouds is dominated by
MHD waves; those cores may be formed, in a few million years, in a 
super--sonic and super--Alfv\'{e}nic flow, only marginally affected by the
magnetic field (model Ad2). The dynamics becomes strongly affected by the magnetic
field only in some very dense regions, on small scales, and preferentially
during the first dynamical time.

We stress that a key theoretical ingredient to the interpretation
of OH Zeeman splitting data is the $B$--$n$ relation. The fact that the 
exponent of the relation is $>0.5$ for more than one dynamical time
is the reason why sub--Alfv\'{e}nic cores can be found in the experiment.

An interesting question is whether the present results, that 
indicate that kinetic energy dominates over magnetic energy on the 
scale of cold clouds, could still be consistent with the well established
rough equipartition of kinetic and magnetic energies on larger scales,
e.g.\ comparable with the thickness of the galactic disc (\cite{Boulares+90}).
Here, we can only offer some preliminary remarks; a more quantitative 
answer requires larger scale and higher resolution numerical simulations, 
that go beyond our current numerical capabilities.  

However, we have made a rather interesting observation already 
in the current simulations.  It turns out that, while the power 
spectra of velocity and magnetic field strength are nearly 
parallel, the spectra of kinetic and magnetic energy are not.  
Because of the intermittency of mass density $\rho$ that develops 
in these supersonic flows, the power spectrum of $\rho^{\frac{1}{2}} 
\vv$ need not be parallel to the power spectrum of $\vv$.
In our super-Alfv{\'e}nic experiments there is a difference of 
slope between the power spectra of magnetic and kinetic energy, 
amounting to about $0.5$, with the kinetic energy spectrum being
more shallow.

Such a difference in slope opens up the interesting possibility
that, over the 2--3 orders of magnitude that separate cold clouds
from large scale disc structures, a difference of about an order
of magnitude develops between the kinetic and magnetic energy 
densities.  This is roughly consistent with the 
differences between kinetic and magnetic energy adopted in our
super-Alfv{\'e}nic models.   More firm conclusion on this issue
must await models with higher Mach number and higher numerical 
resolution.

\section{Conclusions}

In this work we have shown that:

\begin{itemize}

\item Both super--Alfv\'{e}nic and equipartition random super--sonic 
flows, are characterized by a decay time--scale approximately equal 
to one dynamical time. One of the original motivations for the 
theoretical models of magnetized clouds with rough equipartition between
kinetic, gravitational, and magnetic energies was the belief that in
such models the decay time--scale could be significantly longer than 
one dynamical time. Since this is now known to be incorrect
(see also MacLow et al. 1998 and Stone, Ostriker \& Gammie 1998), 
equipartition models of molecular clouds have lost their original motivation. 

\item Random super--sonic and super--Alfv\'{e}nic motions produce 
a very intermittent probability distribution of magnetic field strength, 
with an exponential tail. Cores with field strength several times in 
excess of the mean field can be formed.

\item 
\begin{sloppypar}
Random super--sonic and super--Alfv\'{e}nic motions also produce 
a very intermittent probability distribution of gas density,
in agreement with stellar extinction measurements, while 
equipartition models fail to reproduce the observational result.
\end{sloppypar}

\item Synthetic spectra calculated from both the super--Alfv\'{e}nic
and the equipartition models are used to study the correlation between
line width and gas column density. The equipartition models fail
again to reproduce qualitatively the growth of line width with
column density, while the super--Alfv\'{e}nic model shows 
a correlation similar to the observed one. 

\item Dense cores, with magnetic pressure larger than thermal pressure,
and velocity dispersions smaller than $v_{A}$, are found as the result of 
the evolution of supersonic and super--Alfv\'{e}nic flows. 

\item A power law statistical $B$--$n$ relation 
is generated by super--sonic and super--Alfv\'{e}nic motions, but not by
MHD waves in the equipartition model. The exponent of the relation is $>0.5$ 
for about one dynamical time, which allows for the existence of cores 
with $v_{A}>C_{S}$.

\item The exponent in the $B$--$n$ relation of the super--Alfv\'{e}nic model
decreases with time, because magnetic field lines are stretched and 
partially aligned with the flow. The statistical importance of the 
magnetic pressure in dense cores thus decreases with time, even in 
the absence of gravity and ambipolar diffusion. After about one dynamical
time, the exponent of the upper envelope of the $B$--$n$ relation,
is approximately equal to 0.4, as in the observations.

\item The scatter in the $B$--$n$ relation of the super--Alfv\'{e}nic model
is also consistent with the Zeeman splitting measurements, while the scatter
in the equipartition model is inconsistent with most Zeeman
splitting upper limits on the field strength, and even with some field detections.

\end{itemize}

In summary, even though the evidence presented in the present paper 
is far from complete, it appears that super--Alfv\'{e}nic
motions produces in a natural way the
same type of statistical properties that characterize cold molecular
clouds, while motions with equipartition of kinetic and magnetic 
energies have properties that are harder to reconcile
with the observations. 

\acknowledgements

This work has been partially supported by the Danish National Research Foundation
through its establishment of the Theoretical Astrophysics Center.

We thank Enrique V\'{a}zquez--Semadeni for a very detailed and very useful 
(second opinion) referee
report, and Prof. L. Mestel and Prof. E. Zweibel for their comments on
the manuscript.


\onecolumn
\clearpage

{\bf Figure and Table captions:} \\

{\bf Table 1:}  Initial parameters of the numerical experiments. \\
${\cal M}_A$: alfv\'{e}nic rms Mach number; \\
$b_0$: magnetic field strength parameter; \\
${\cal M}$: rms Mach number; \\
ambi: ambipolar diffusion parameter; \\

{\bf Table 2:} Initial physical parameters of the models, obtained by scaling
the numerical parameters of the experiments (Table 1) with the 
Larson's relations (Eqs. {7} and (8)). \\

\begin{figure}
\centerline{\epsfxsize=15cm \epsfbox{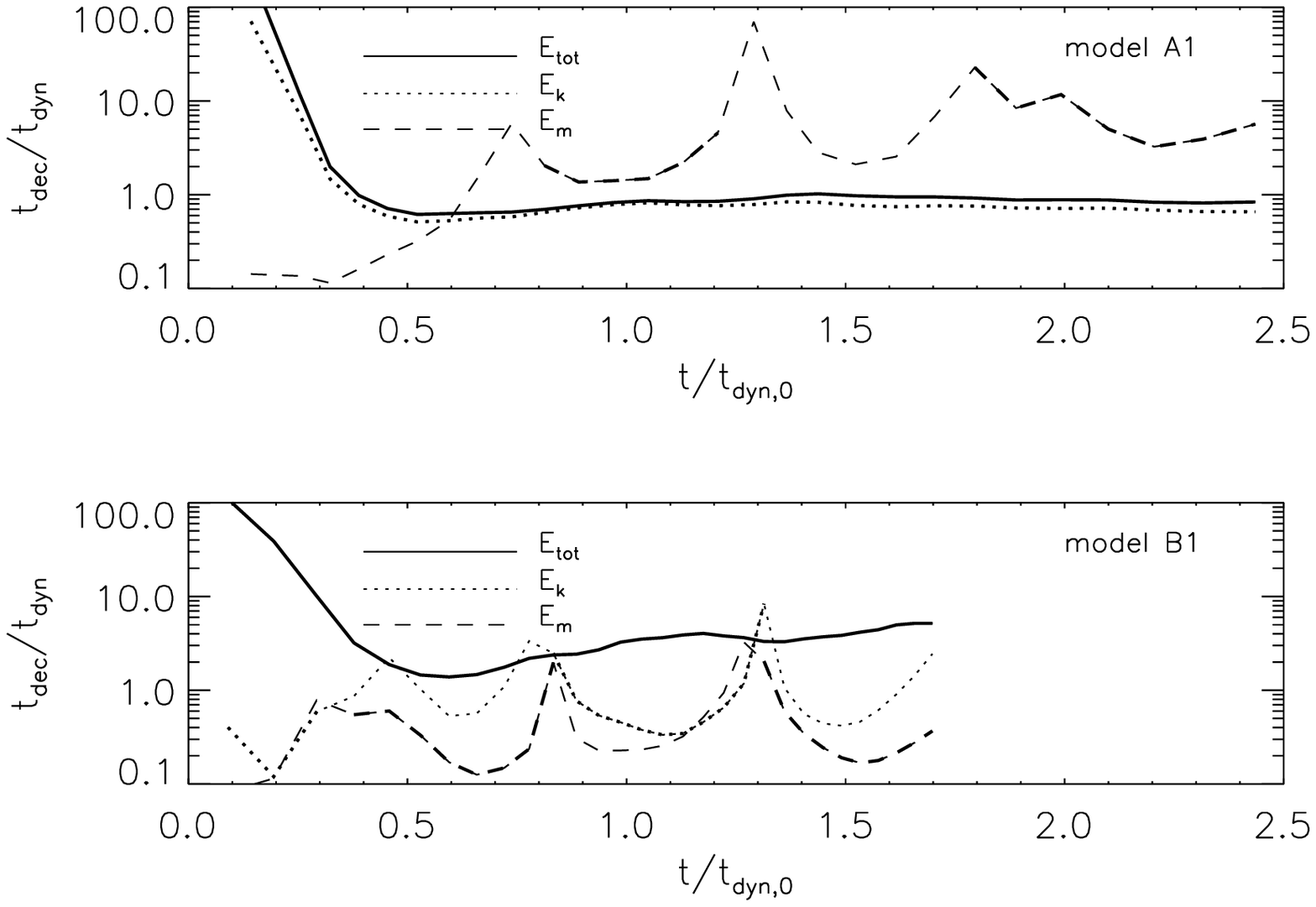}}
\caption[]{}
\label{fig1}
\end{figure}

{\bf Fig.\ \ref{fig1}:}  Time evolution of the ratio of the energy decaying time 
divided by the instantaneous dynamical time. Separate plots are for the
total energy $E_{tot}$, the magnetic energy $E_m$, and the kinetic energy
$E_k$.\\

\begin{figure}
\centerline{\epsfxsize=15cm \epsfbox{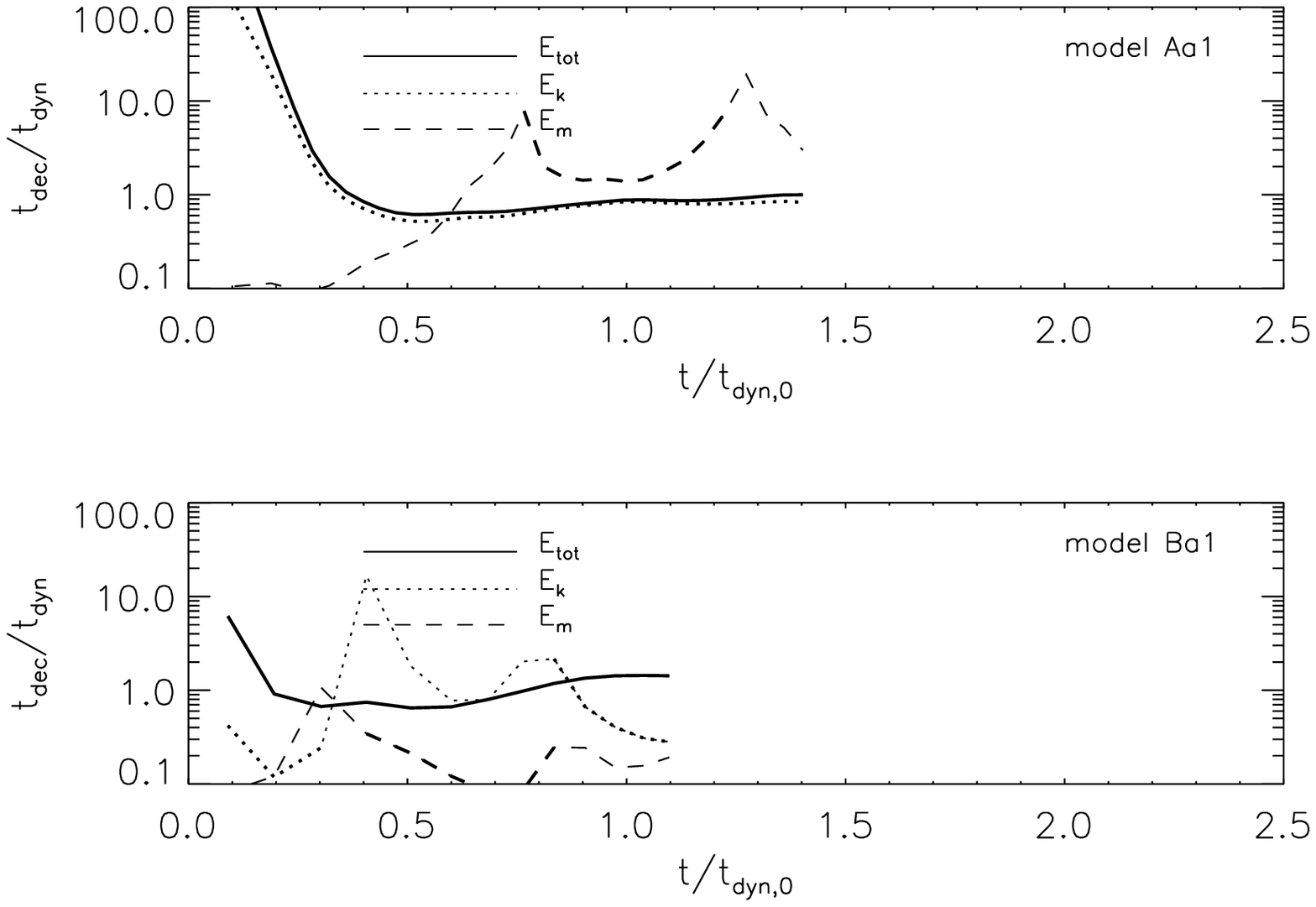}}
\caption[]{}
\label{fig2}
\end{figure}

{\bf Fig.\ \ref{fig2}:} Same as in Fig.~\ref{fig1}, but for models with ambipolar 
diffusion.\\

\begin{figure}
\centerline{\epsfxsize=15cm \epsfbox{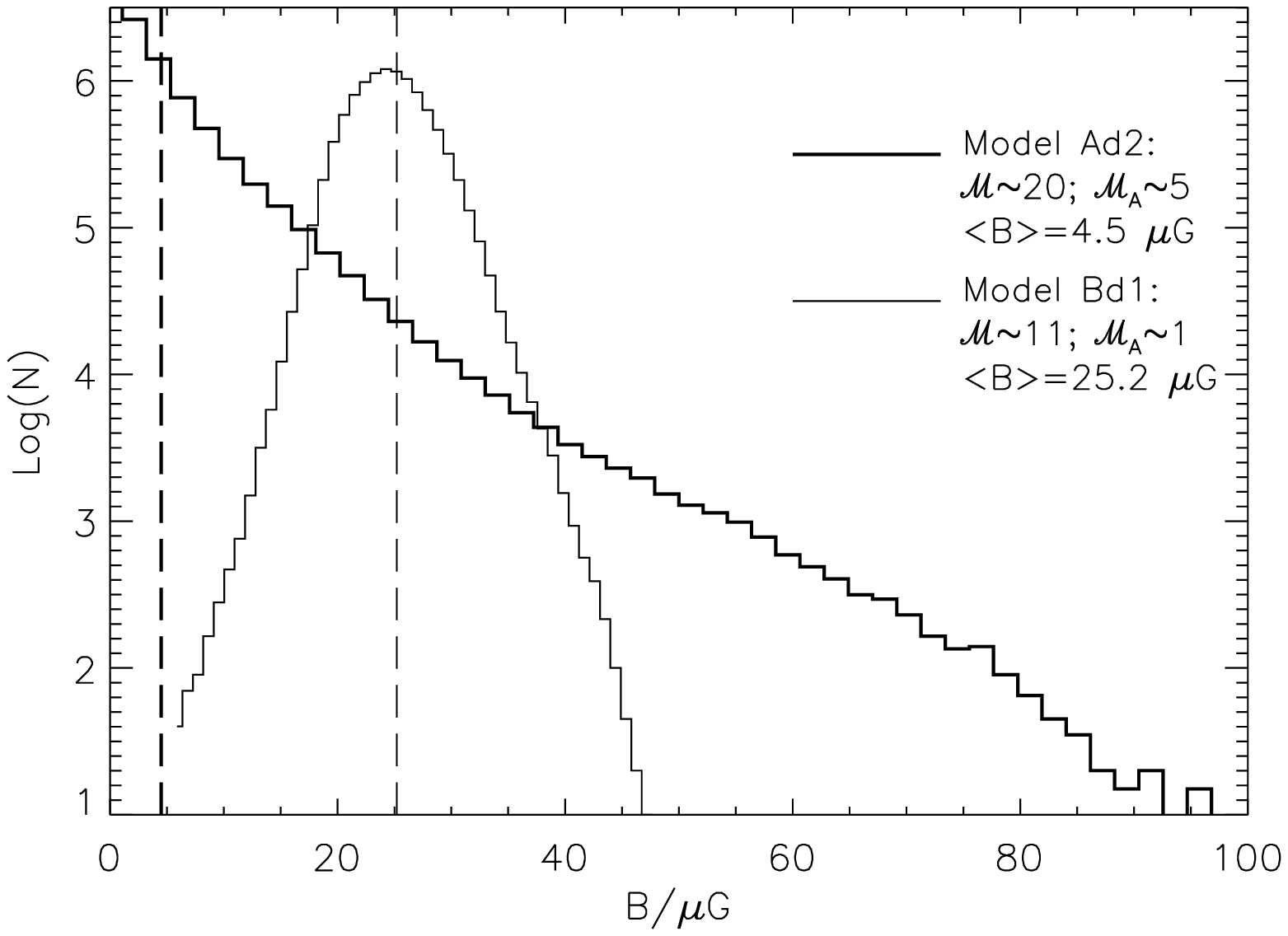}}
\caption[]{}
\label{fig3}
\end{figure}

{\bf Fig.\ \ref{fig3}:} Distribution of magnetic field strength in the 
super--Alfv\'{e}nic model Ad2 (thick line), and in the 
equipartition model Bd1 (thin lines). The dashed vertical lines mark the 
mean field strength in the two models. \\

\begin{figure}
\centerline{\epsfxsize=11cm \epsfbox{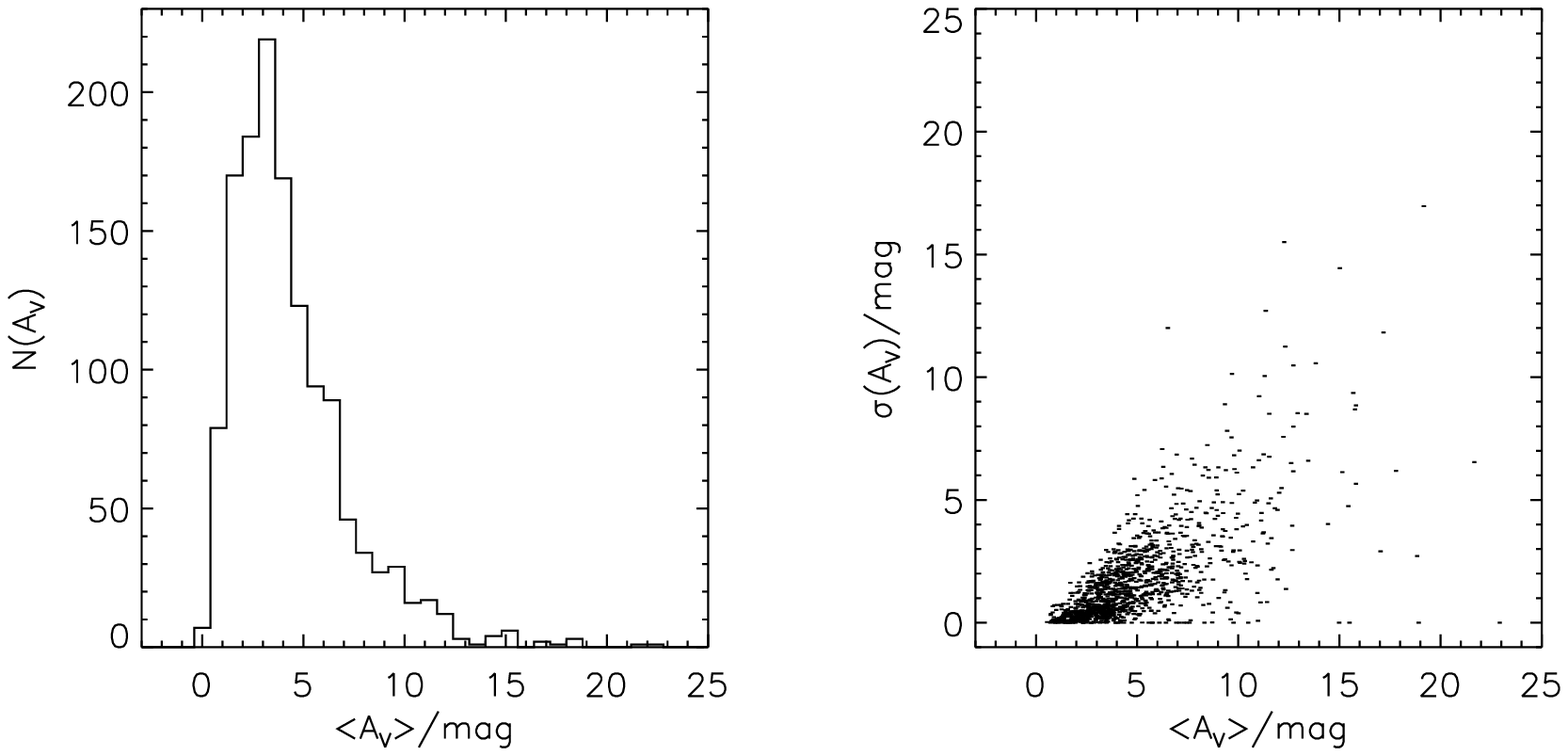}}
\centerline{\epsfxsize=11cm \epsfbox{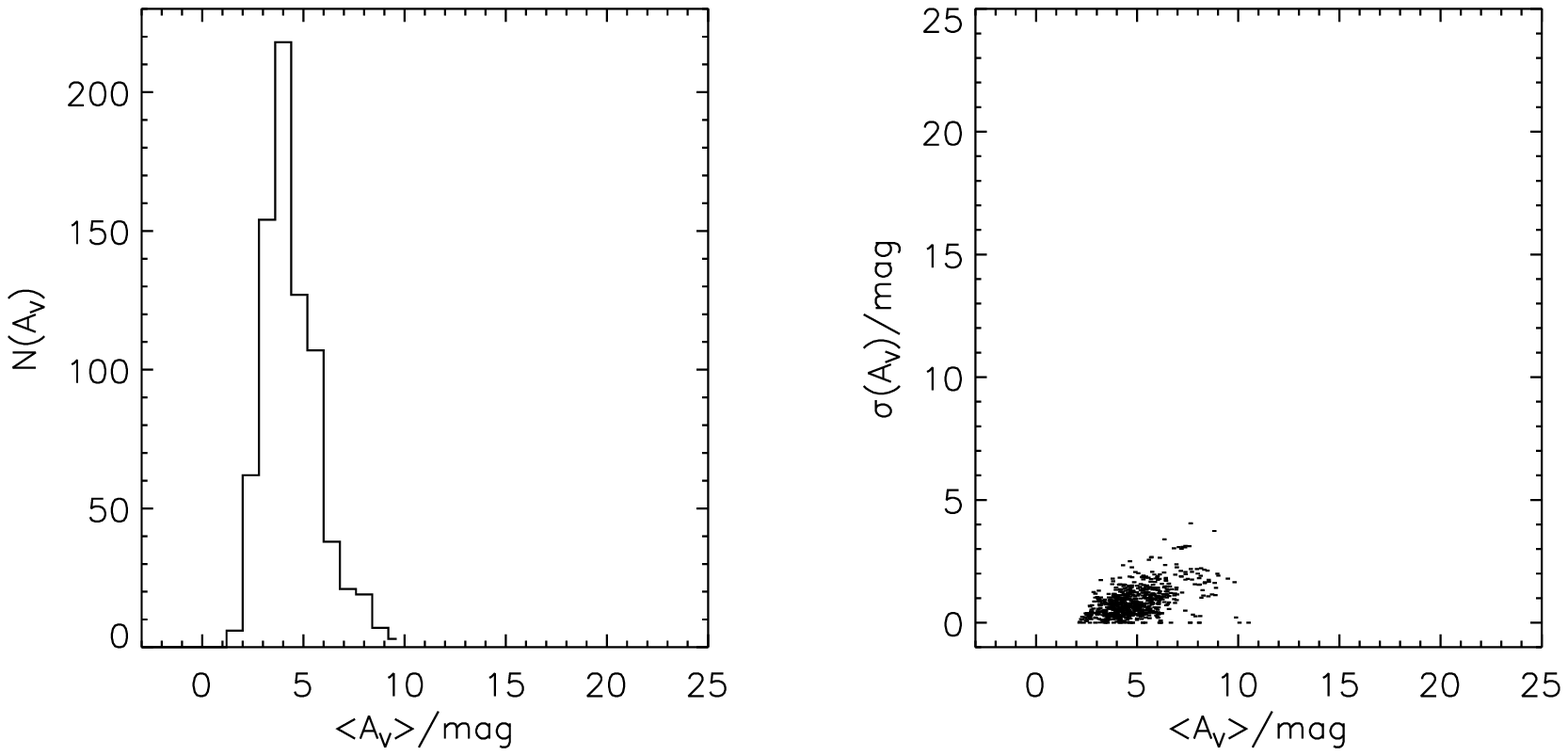}}
\centerline{\epsfxsize=11cm \epsfbox{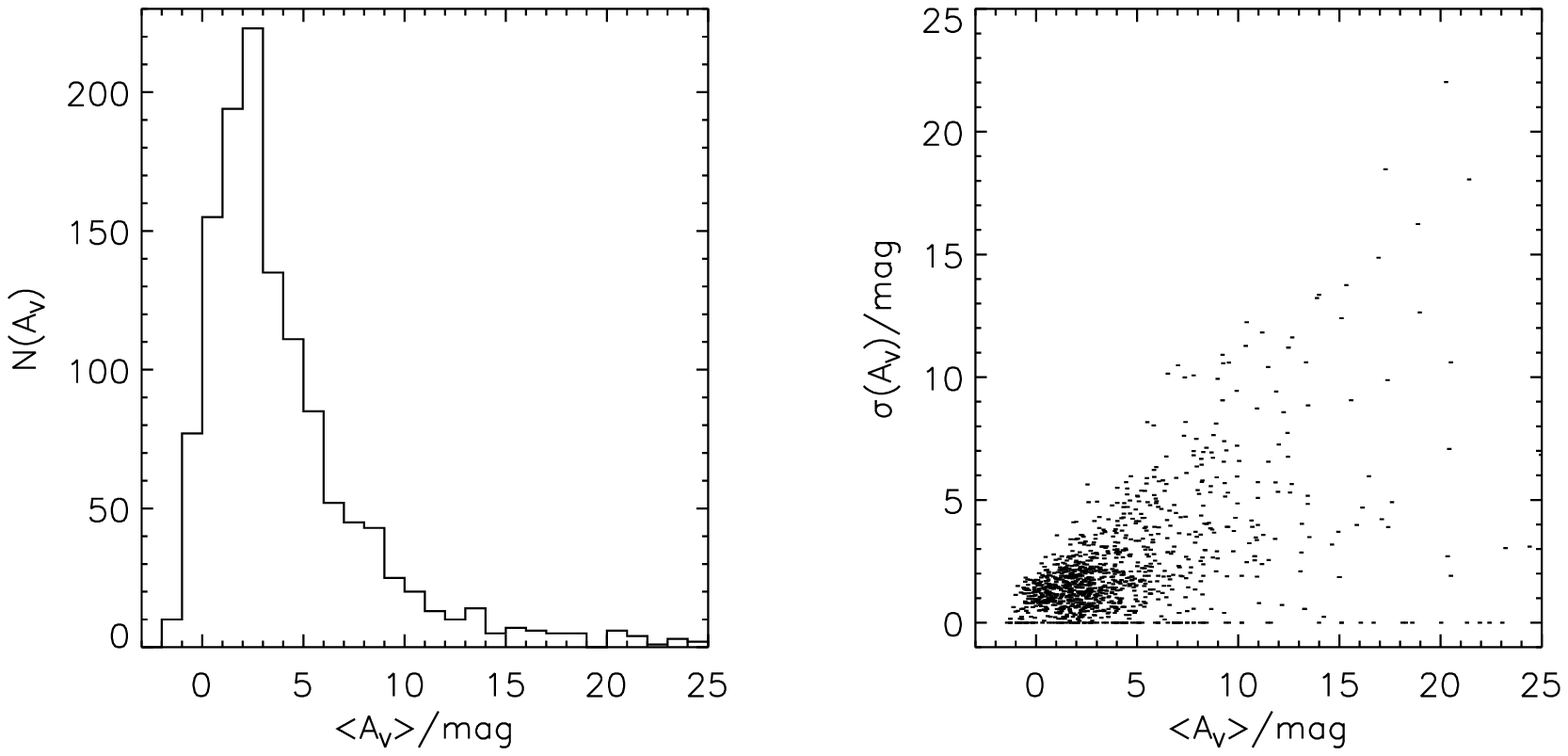}}
\caption[]{}
\label{fig4}
\end{figure}

{\bf Fig.\ \ref{fig4}:} Histograms of extinction (left panels) and plots
of dispersion of extinction in cells versus the mean cell extinction
(right panels). The top panels show the result from the super--Alfv\'{e}nic
model Ad2, while the middle panels are from the equipartition model Bd1.
The bottom panels are \citetext{Lada+94} observational data for the 
cloud IC5146. \\

\begin{figure}
\centerline{\epsfxsize=15cm \epsfbox{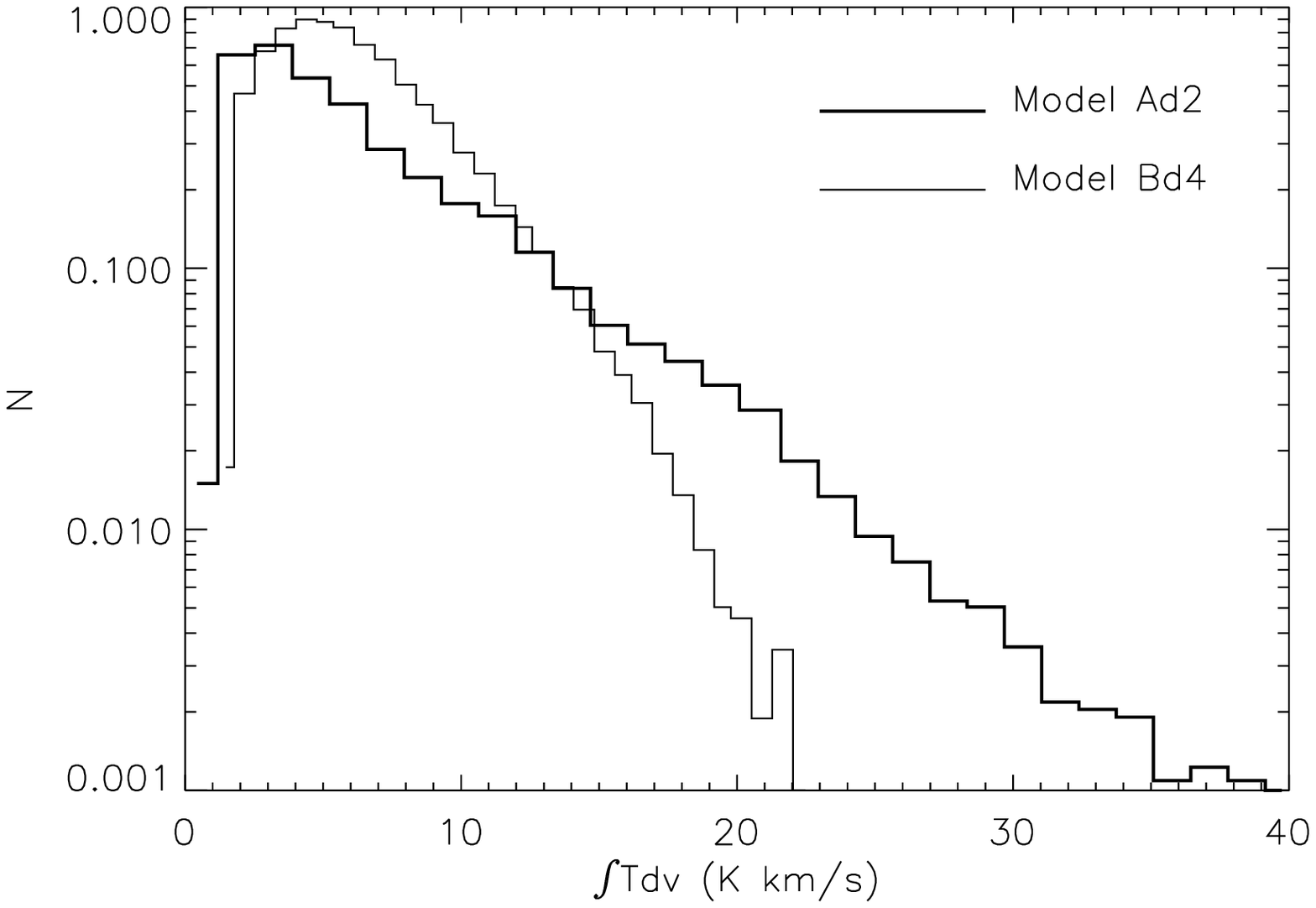}}
\caption[]{}
\label{fig5}
\end{figure}

{\bf Fig.\ \ref{fig5}:} Histograms of integrated antenna temperature of
synthetic \Jtco\ spectra, from the super--Alfv\'{e}nic model Ad2
(thick line) and for the equipartition model Bd1 (thin line).
Model Ad2 shows a very intermittent approximately exponential
distribution. \\

\begin{figure}
\centerline{\epsfxsize=15cm \epsfbox{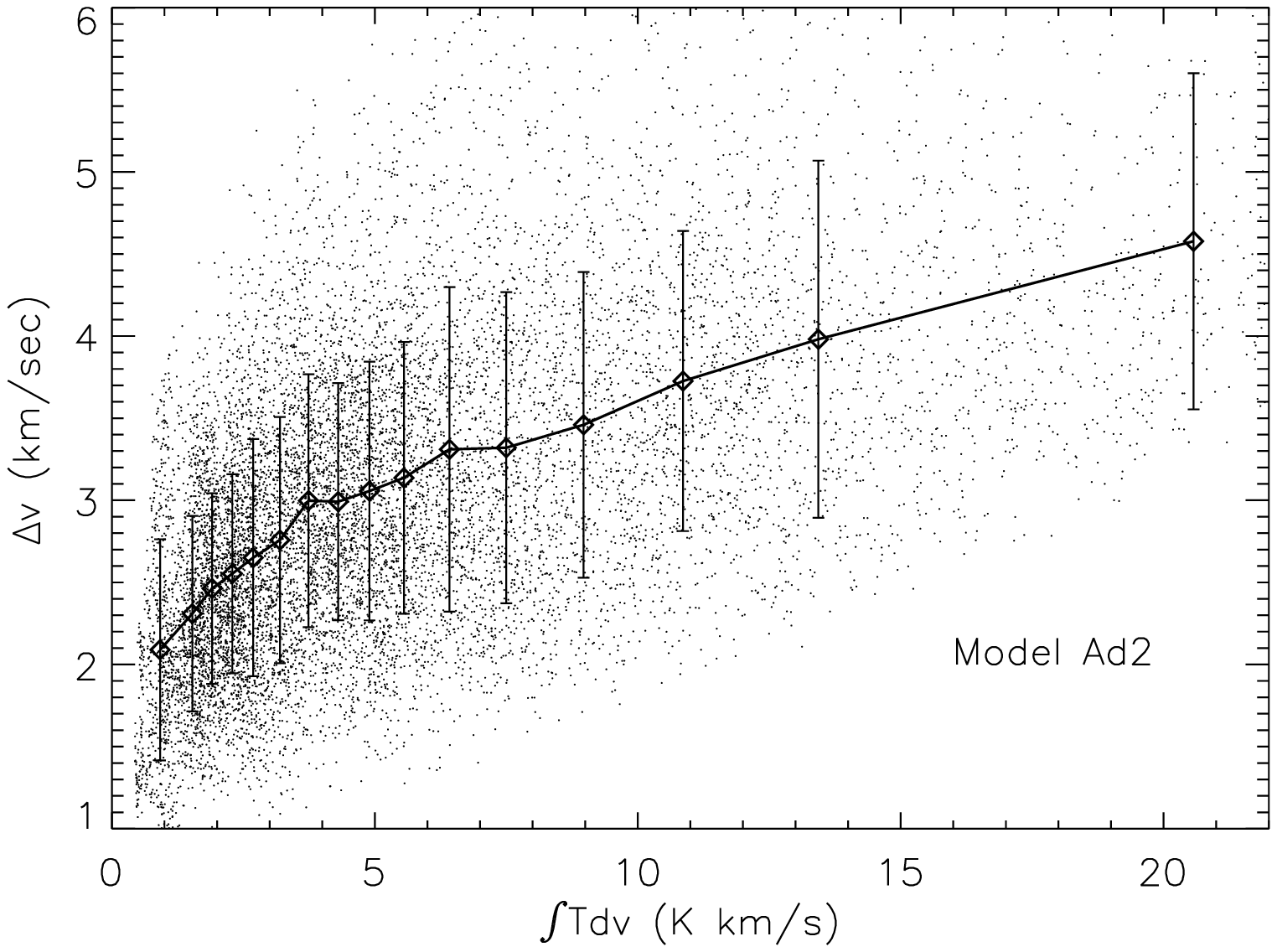}}
\centerline{\epsfxsize=15cm \epsfbox{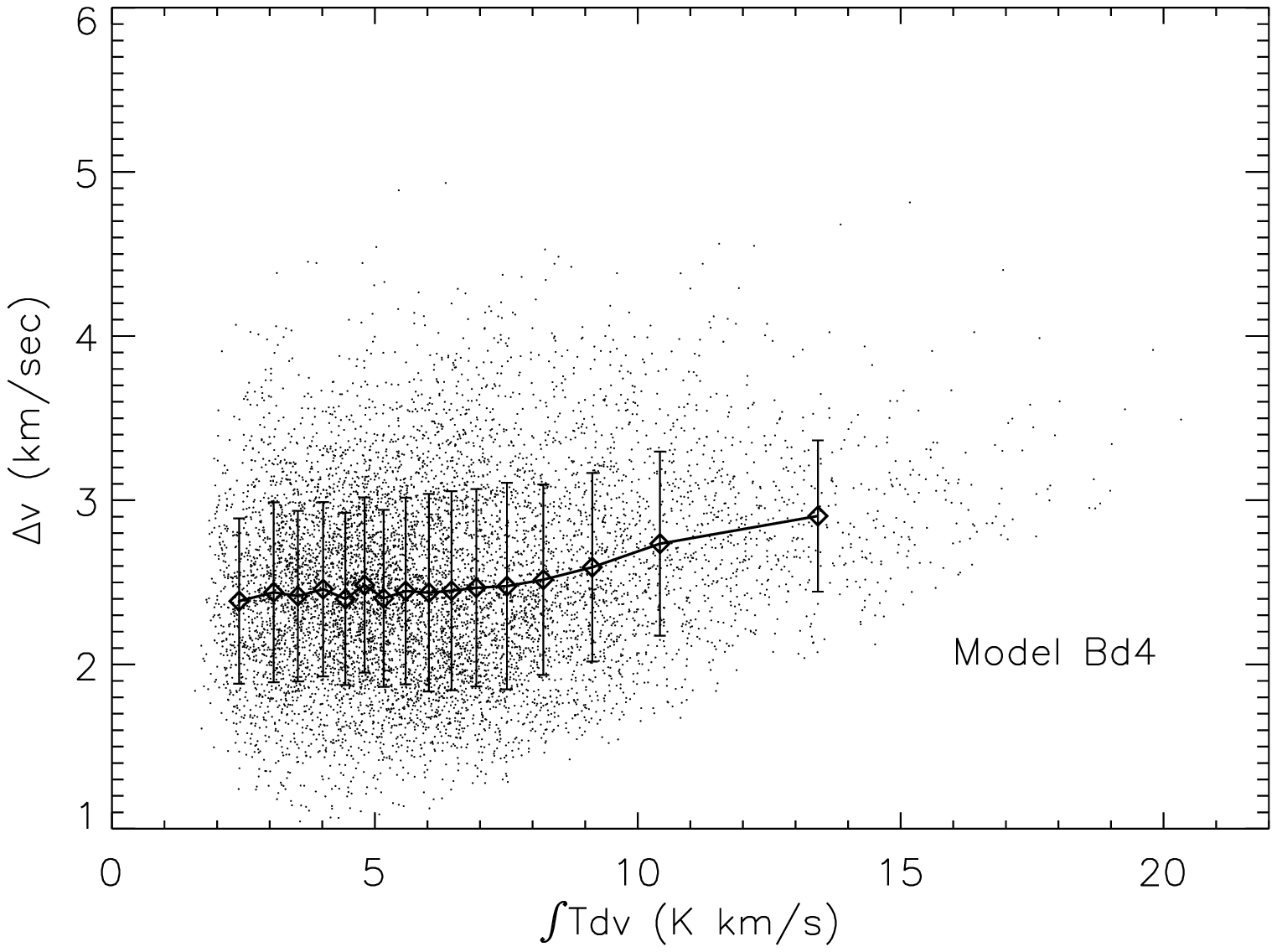}}
\caption[]{}
\label{fig6}
\end{figure}

{\bf Fig.\ \ref{fig6}:} \Jtco\ equivalent width versus \Jtco\ integrated
antenna temperature, from the super--Alfv\'{e}nic model Ad2
(upper panel) and for the equipartition model Bd1 (lower panel). \\

\begin{figure}
\centerline{\epsfxsize=15cm \epsfbox{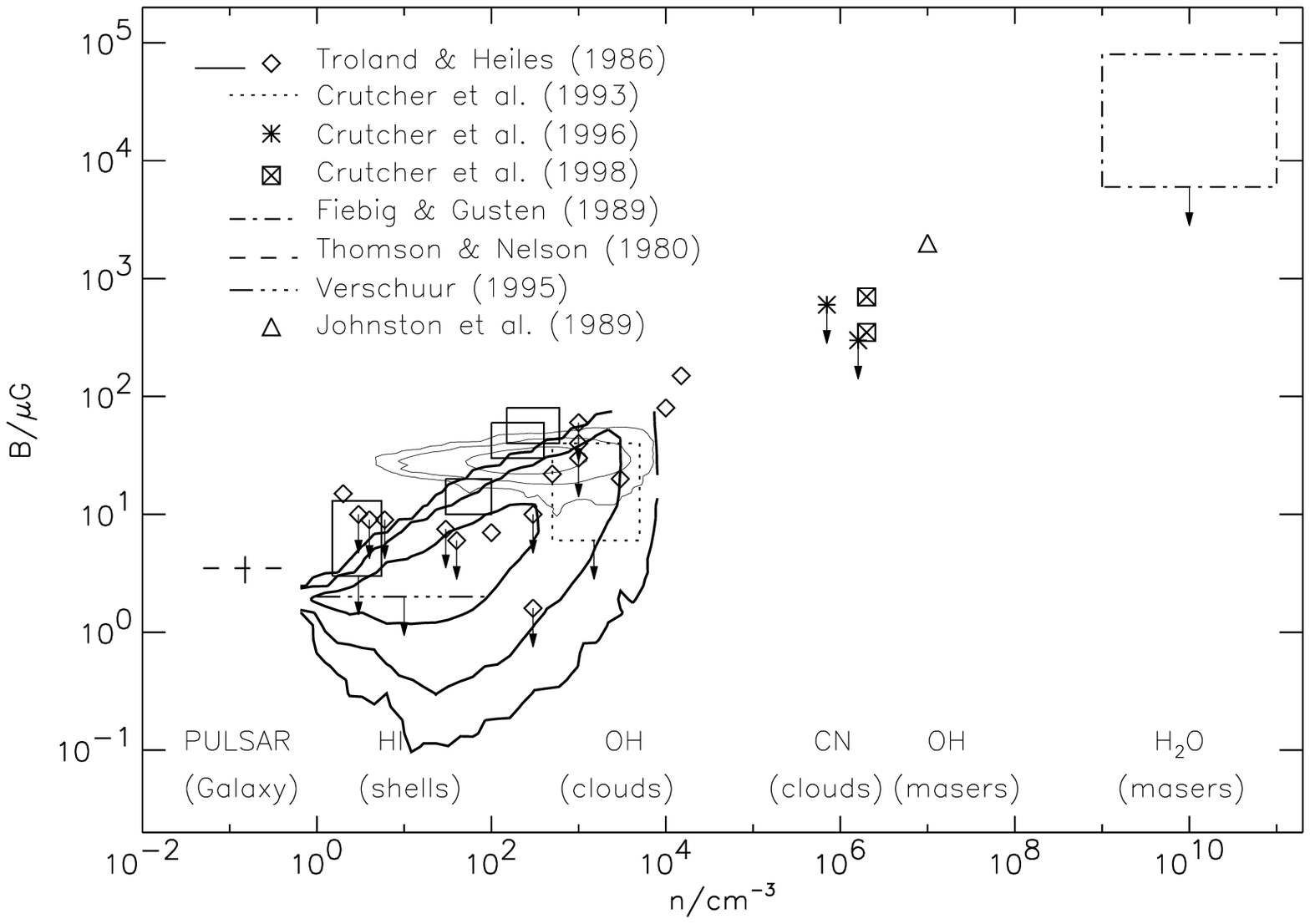}}
\caption[]{}
\label{fig7}
\end{figure}

{\bf Fig.\ \ref{fig7}:} The $B-n$ relation: observations and theoretical 
models. The thick contour lines are from the the super--Alfv\'{e}nic
model Ad2, and the thin contour lines from the equipartition model Bd1. \\

\begin{figure}
\centerline{
    \epsfxsize=8cm\epsfbox{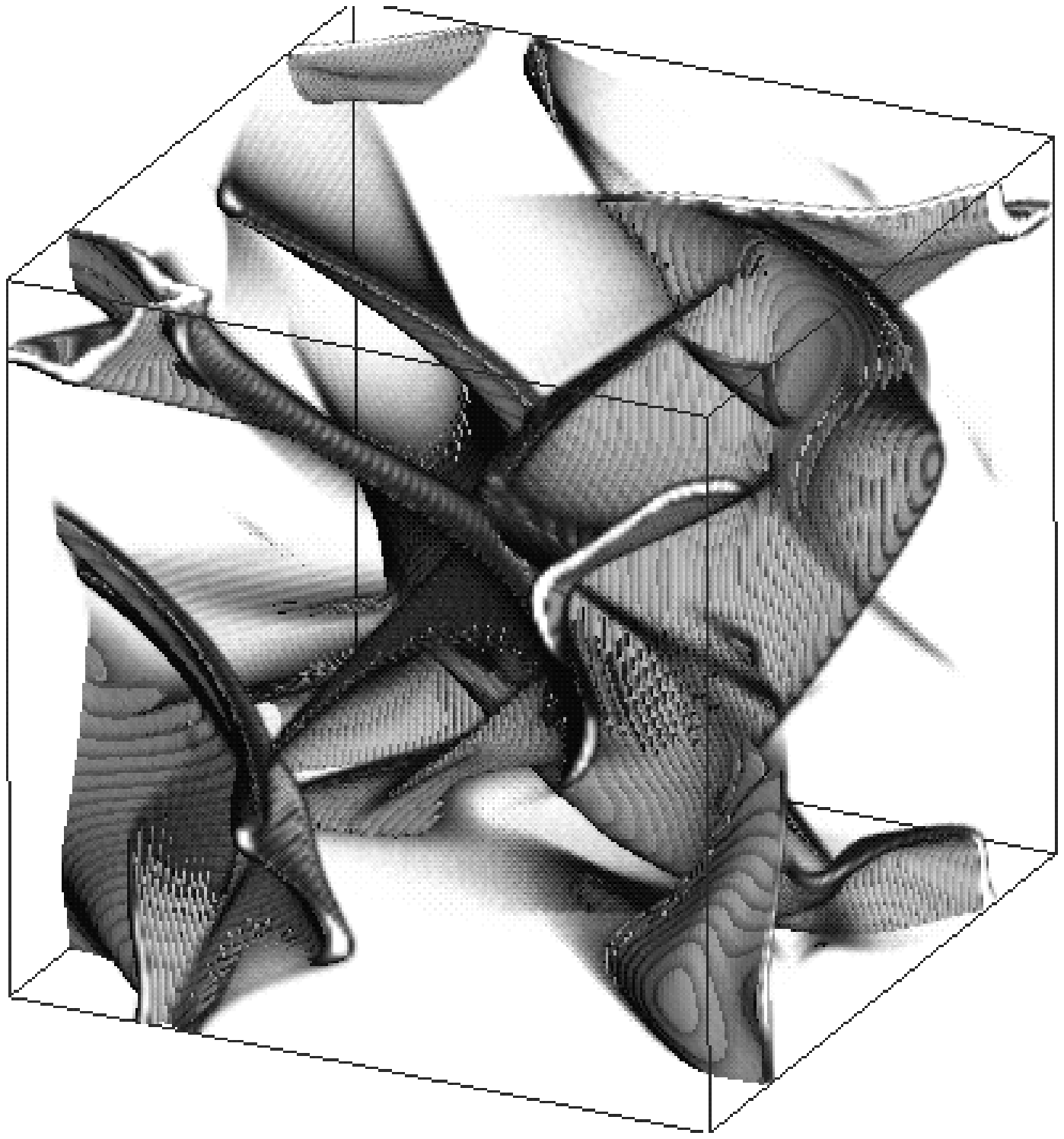}
    \epsfxsize=8cm\epsfbox{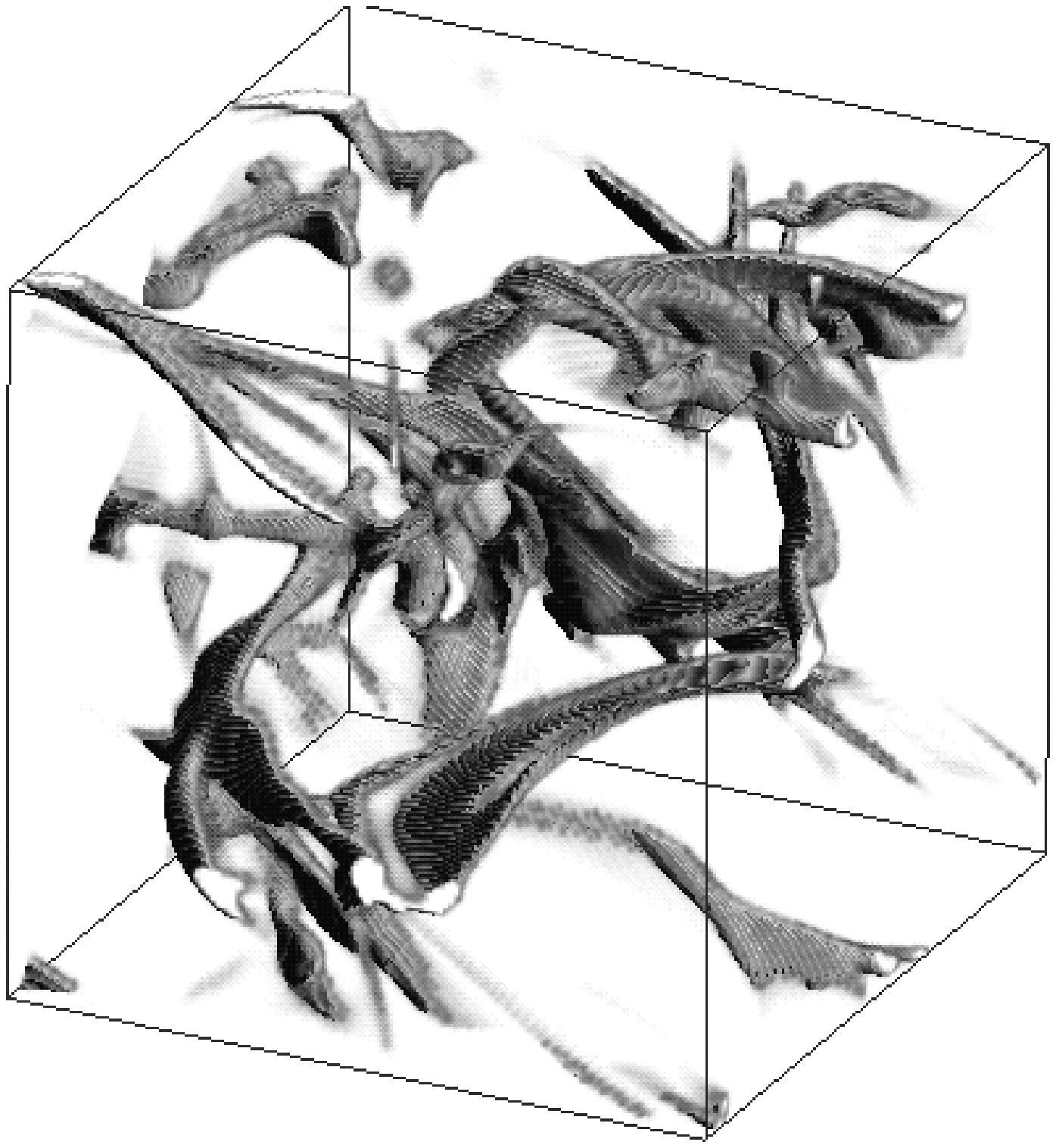}
}
\caption[]{}
\label{fig8a}
\end{figure}

{\bf Fig.\ \ref{fig8a}:} Three dimensional visualizations of the 
density field in the super--Alfv\'{e}nic model Ad2, obtained by 
the superposition of voxel projection and density isosurfaces.
The left panel corresponds to $t=0.5t_{dyn,0}$, and the right
panel $t=1.5t_{dyn,0}$. \\

\begin{figure}
\centerline{
    \epsfxsize=8cm\epsfbox{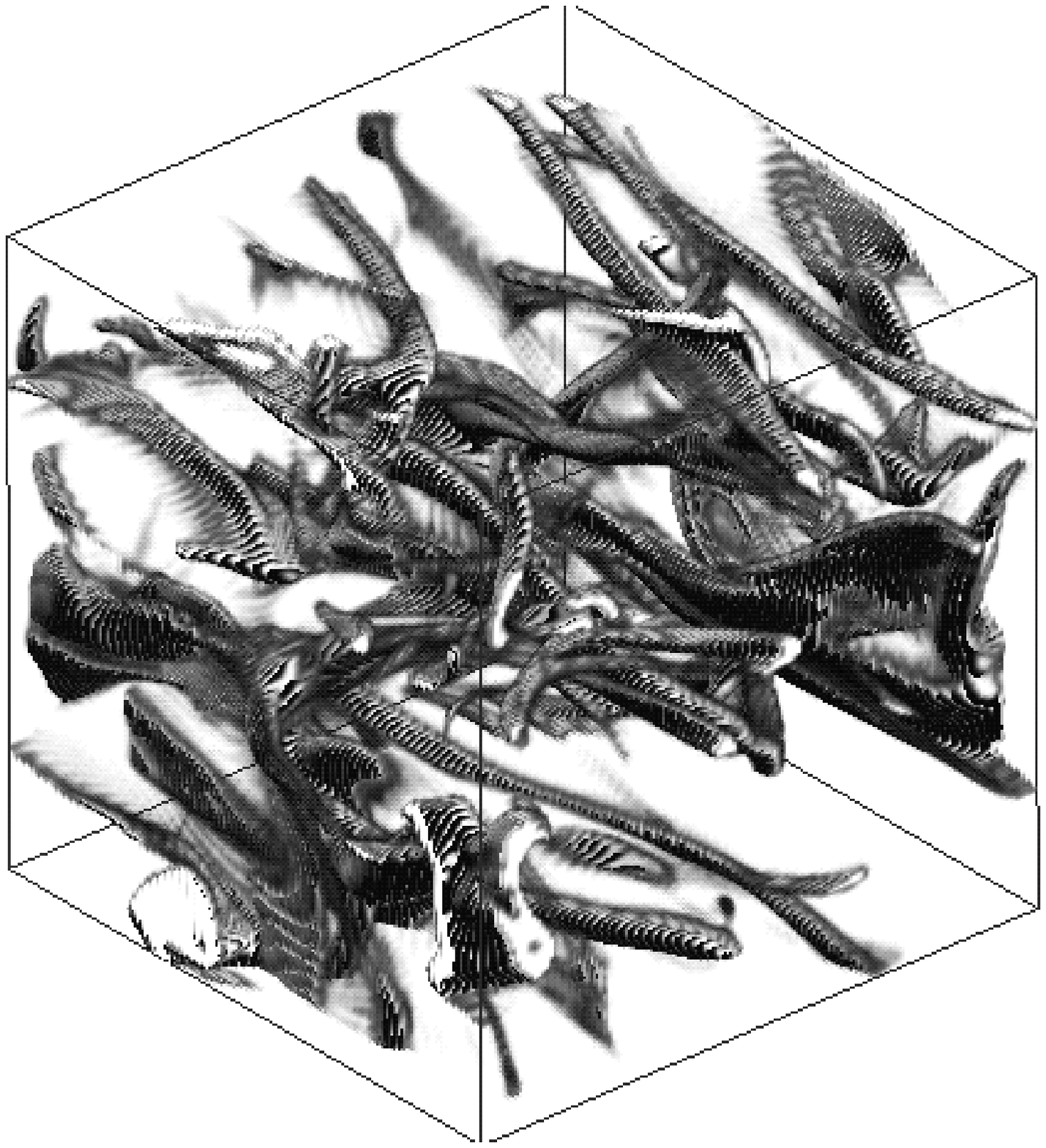}
    \epsfxsize=8cm\epsfbox{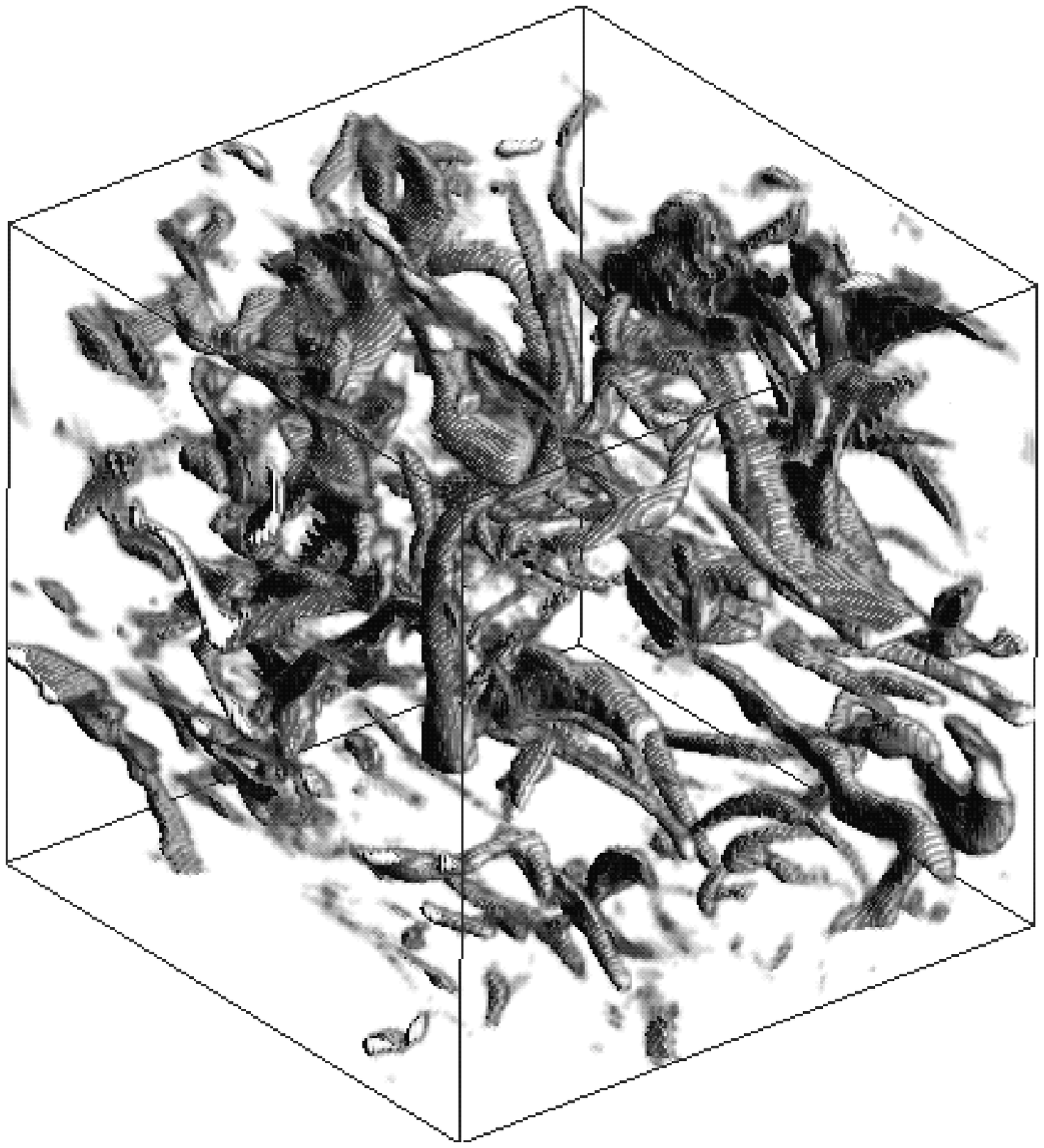}
}
\caption[]{}
\label{fig8b}
\end{figure}

{\bf Fig.\ \ref{fig8b}:} Same as in Fig.~\ref{fig8a} but for the 
equipartition model Bd1. The left panel corresponds to 
$t=1.3t_{dyn,0}$, and the right panel $t=2.8t_{dyn,0}$. \\

\begin{figure}
\begin{center}
  \begin{minipage}{8cm}
    \centerline{
    \epsfxsize=8cm
    \epsfbox{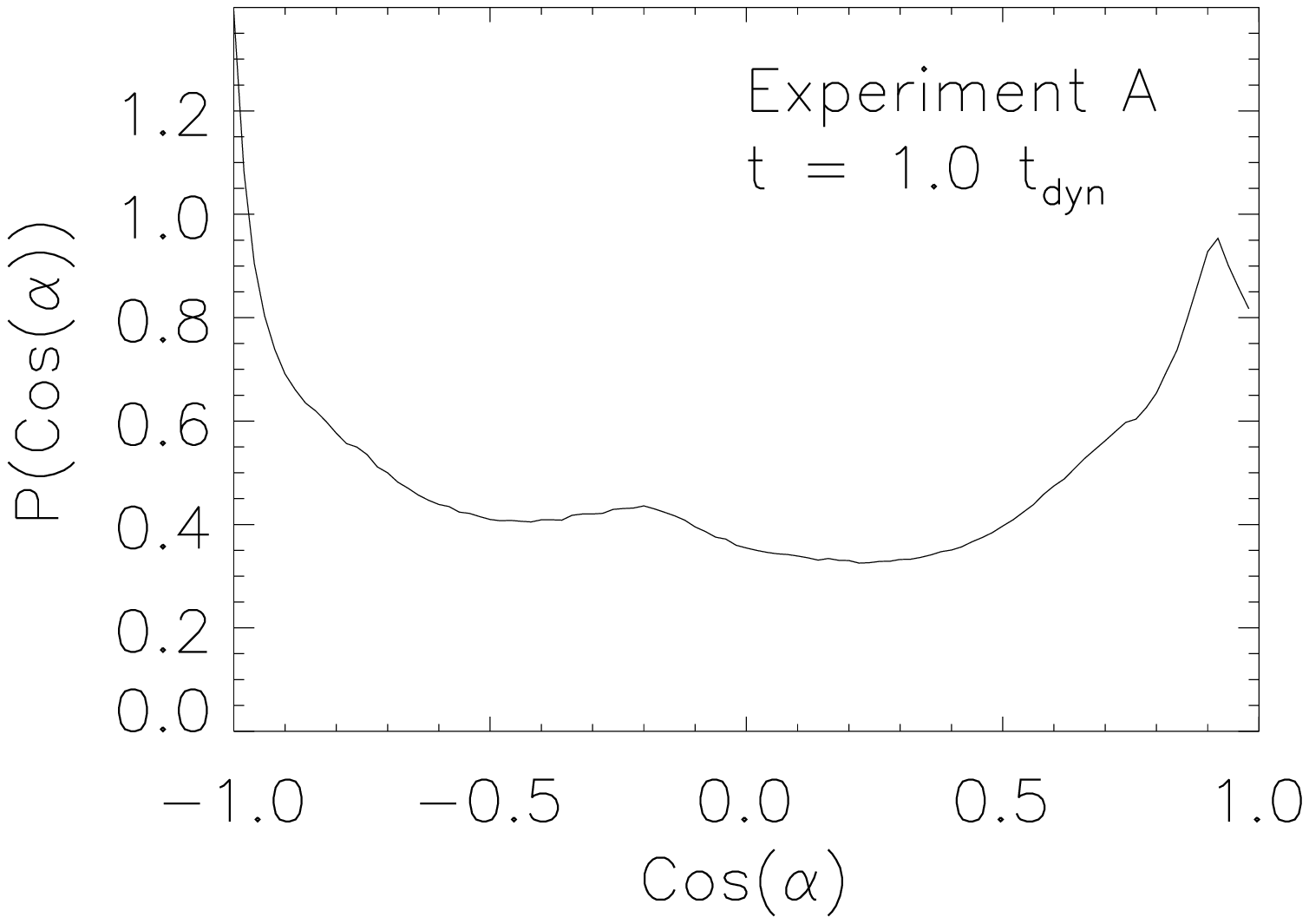}
    }
  \end{minipage}
  \begin{minipage}{8cm}
    \centerline{
    \epsfxsize=8cm
    \epsfbox{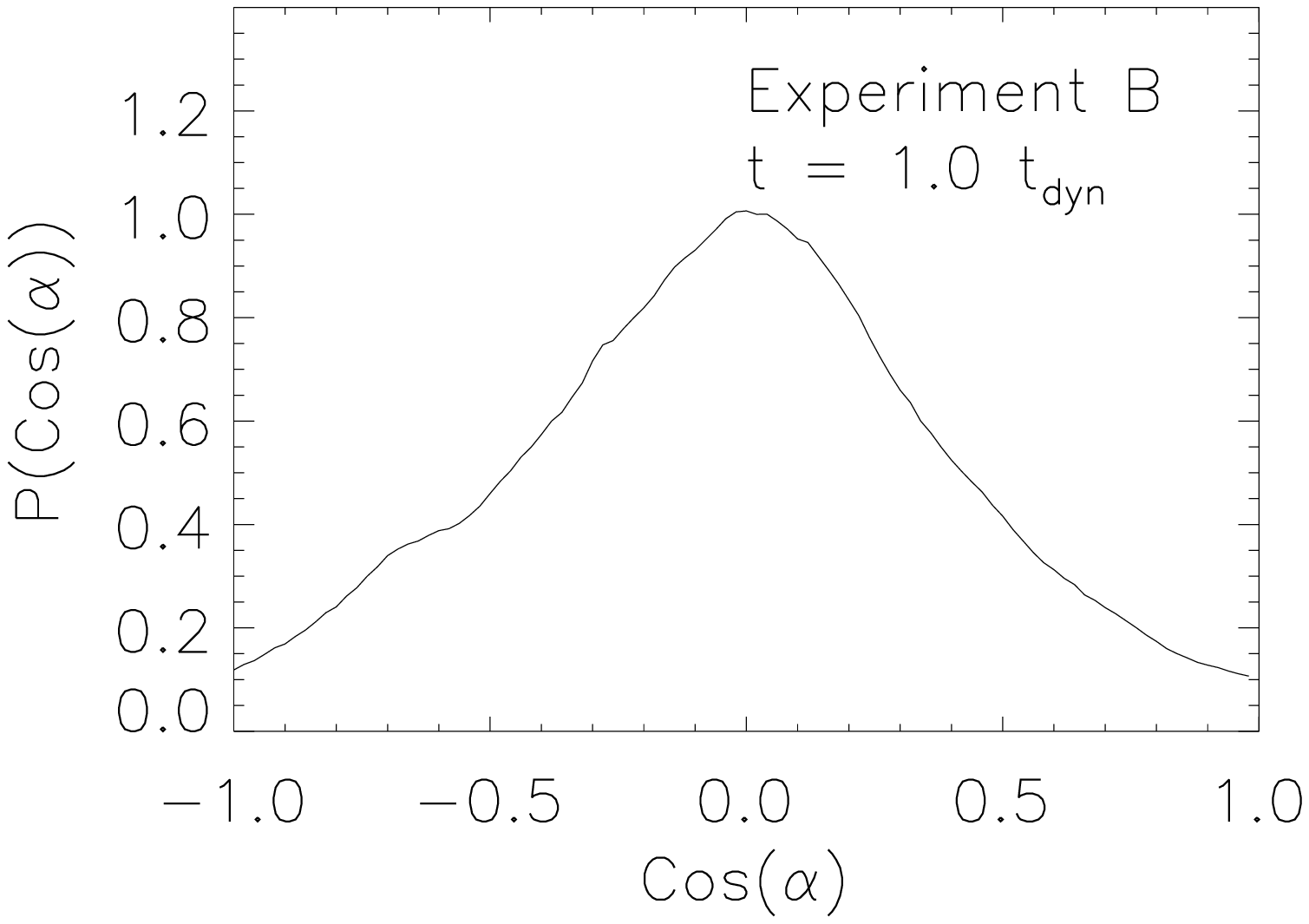}
    }
  \end{minipage}
\caption[]{}
\label{fig9}
\end{center}
\end{figure}

{\bf Fig.\ \ref{fig9}:} Histograms of the cosine of the angle between ${\bf v}$ 
and ${\bf B}$. In experiment A1 (left panel) there is a partial alignment, 
while in experiment B1 (right panel) the two fields are mainly perpendicular 
to each other. \\

\clearpage
\begin{table}
\begin{tabular}{|c|c|c|c|c|c|c|c|c|}
\hline
 Model name      & ${\cal M}_A$     &  $b_0$  & $\beta$ & ${\cal M}$ &  ambi   &  snapshots   &   Remarks   &  File name     \\ 
\hline
   {\bf A1}      &   {\bf 10.6}       &   0.5   &  5.33 &  5.3       &   0.0   &  31      &   decaying  &  run\_1    \\ 
   {\bf Aa1}     &   {\bf 10.6}       &   0.5   &  5.33 &  5.3       &   0.25  &  31      &   decaying  &  run\_11   \\ 
   {\bf Aa2}     &   {\bf 10.6}       &   0.5   &  5.33 &  5.3       &   0.50  &  30      &   decaying  &  run\_8    \\ 
   {\bf B1}      &    {\bf 0.8}       &   5.0   &  0.05 &  3.8       &   0.0   &  31      &   decaying  &  run\_3    \\ 
   {\bf B2}      &    {\bf 1.1}       &   5.0   &  0.05 &  5.4       &   0.0   &  11      &   decaying  &  run\_16   \\ 
   {\bf B3}      &    {\bf 2.7}       &   2.0   &  0.33 &  5.4       &   0.0   &  21      &   decaying  &  run\_17   \\ 
   {\bf Ba1}     &    {\bf 0.8}       &   5.0   &  0.05 &  3.8       &   0.50  &  14      &   decaying  &  run\_7    \\ 
   {\bf Ba2}     &    {\bf 1.3}       &   3.0   &  0.15 &  3.8       &   0.50  &  20      &   decaying  &  run\_12   \\ 
\hline
   {\bf Ad1}     &   {\bf 10.6}       &   1.0   &  1.33 & 10.6       &   0.0   &  30      &   driven    &  run\_18   \\ 
   {\bf Ad2}     &   {\bf 28.3}       &   1.0   &  1.33 & 28.3       &   0.0   &  31      &   driven    &  run\_19   \\ 
   {\bf Bd1}     &    {\bf 0.9}       &  10.0   &  0.01 &  8.8       &   0.0   &  23      &   driven    &  run\_21   \\ 
   {\bf Bd2}     &    {\bf 1.2}       &   3.6   &  0.10 &  6.8       &   0.0   &  31      &   driven    &  run\_23   \\ 
   {\bf Bd3}     &    {\bf 1.5}       &   7.2   &  0.03 & 10.7       &   0.0   &  30      &   driven    &  run\_22   \\ 
   {\bf Bd4}     &    {\bf 2.0}       &   7.0   &  0.03 & 14.2       &   0.0   &  15      &   driven    &  run\_20   \\ 
\hline
\end{tabular}
\caption{}
\end{table}

\clearpage
\begin{table}
\begin{tabular}{|c|c|c|c|c|c|}
\hline
  Model name & $\sigma_v$/(km/s) &  $l$/pc &  $\tau_{dyn}$/(10$^6$yr)  &  $\langle n \rangle$/(100~H$_2$~cm$^{-3}$) &  $B_0$/$\mu$G   \\ 
\hline
     A1      &     1.3           &   1.8   &         1.3               &         11                 &  2.6          \\ 
     Aa1     &     1.3           &   1.8   &         1.3               &         11                 &  2.6          \\ 
     Aa2     &     1.3           &   1.8   &         1.3               &         11                 &  2.6          \\ 
     B1      &     0.9           &   0.9   &         0.9               &         22                 &  36           \\ 
     B2      &     1.3           &   1.8   &         1.3               &         11                 &  26           \\ 
     B3      &     1.3           &   1.8   &         1.3               &         11                 &  10           \\ 
     Ba1     &     0.9           &   0.9   &         0.9               &         22                 &  36           \\ 
     Ba2     &     0.9           &   0.9   &         0.9               &         22                 &  22           \\ 
\hline
     Ad1     &     2.6           &   7.0   &         2.6               &         2.8                &  2.6          \\ 
     Ad2     &     7.1           &    50   &         7.1               &         0.4                &  1.0          \\ 
     Bd1     &     2.2           &   4.8   &         2.2               &         4.1                &  31           \\ 
     Bd2     &     1.7           &   2.9   &         1.7               &         6.9                &  15           \\ 
     Bd3     &     2.7           &   7.1   &         2.7               &         2.8                &  19           \\ 
     Bd4     &     3.5           &    13   &         3.5               &         1.6                &  14           \\ 
\hline
\end{tabular}
\caption{}
\end{table}

\nonstopmode

\clearpage

\clearpage

\clearpage

\clearpage

\clearpage

\clearpage

\clearpage

\clearpage

\clearpage

\clearpage

\end{document}